\documentstyle[11pt,epsf,psfig]{article}

\def\fnote#1#2{\begingroup\def\thefootnote{#1}\footnote{#2}\addtocounter
{footnote}{-1}\endgroup}
\newcommand{\AmS}{{\protect\the\textfont2
 A\kern-.1667em\lower.5ex\hbox{M}\kern-.125emS}}
\def\inbar{\vrule height1.5ex width.4pt depth0pt}
\def\IB{\relax{\rm I\kern-.18em B}}
\def\IC{\relax\,\hbox{$\inbar\kern-.3em{\rm C}$}}
\def\ID{\relax{\rm I\kern-.18em D}}
\def\IE{\relax{\rm I\kern-.18em E}}
\def\IF{\relax{\rm I\kern-.18em F}}
\def\IG{\relax\,\hbox{$\inbar\kern-.3em{\rm G}$}}
\def\IH{\relax{\rm I\kern-.18em H}}
\def\II{\relax{\rm I\kern-.18em I}}
\def\IK{\relax{\rm I\kern-.18em K}}
\def\IL{\relax{\rm I\kern-.18em L}}
\def\IM{\relax{\rm I\kern-.18em M}}
\def\IN{\relax{\rm I\kern-.18em N}}
\def\IO{\relax\,\hbox{$\inbar\kern-.3em{\rm O}$}}
\def\IP{\relax{\rm I\kern-.18em P}}
\def\IQ{\relax\,\hbox{$\inbar\kern-.3em{\rm Q}$}}
\def\IR{\relax{\rm I\kern-.18em R}}
\def\ZZ{\relax{\sf Z\kern-.4em Z}}
\def\fnote#1#2{\begingroup\def\thefootnote{#1}\footnote{#2}\addtocounter
{footnote}{-1}\endgroup}
\def\beq{\begin{equation}}
\def\eeq{\end{equation}}
\def\bea{\begin{eqnarray}}
\def\eea{\end{eqnarray}}
\def\lleq#1{\label{#1}\eeq}
\def\llea#1{\label{#1}\eea}
\let\nn=\nonumber
\def\tabroom{\hbox to0pt{\phantom{\huge A}\hss}}
\def\notin{\ \hbox{{$\in$}\kern-.51em\hbox{/}}}

\def\a{\alpha}   
   \def\l{\lambda}
   \def\si{\sigma}
  \def\th{\theta}
\def\cA{{\cal A}}  
\def\cD{{\cal D}}
\def\cF{{\cal F}}  \def\cH{{\cal H}}
\def\cI{{\cal I}}

  \def\cP{{\cal P}}

 \def\td{{\tilde d}} \def\tk{{\tilde k}}

 \def\wtW{{\widetilde W}}

\def\dhat{{\hat{d}}} \def\khat{{\hat{k}}}

\def\ba{\bar a}   \def\bd{\bar d}
   \def\bk{\bar k}

  \def\bp{\bar p} 
 \def\bt{\bar t}  
 \def\bz{\bar z}
\def\bA{\bar A}

\def\bcD{\bar \cD} \def\bcF{\bar \cF}

 \def\bphi{{\bar \phi}} 

\def\bth{{\bar \theta}} 
\def\bPhi{{\bar \Phi}}

\def\ra{{\rightarrow}}
\def\lra{\longrightarrow}
\def\lolra{\longleftrightarrow}

\def\del{\partial}

\def\hbar{\bar h}

 \def\II{{\bf II}}

\def\fth{{{\rm F}_{12}}}

\def\het{{\rm Het}}
\def\cyfour{{{\rm CY}_4}}
\def\cythree{{{\rm CY}_3}}

\begin{document}

\hfill {hep-th/9812195}
\vskip .01truein
\hfill {BONN--TH--98--13}
\vskip 1truein

\centerline{\bf {\Large Landau--Ginzburg Vacua of String,
 M- and F-Theory at $c=12$}}

\vskip .4truein

\centerline{ {\sc Monika Lynker$^1$}\fnote{\star}{e--mail:\
 mlynker@rust.iusb.edu},~
 {\sc Rolf Schimmrigk$^{1,2}$}\fnote{\dagger}{e--mail:\
 netah@rust.iusb.edu}
 \ \ and \
 {\sc Andreas Wi\ss kirchen$^2$}\fnote{*}{e--mail:\
 wisskirc@avzw02.physik.uni--bonn.de}
 }
\bigskip
\smallskip
\centerline{${}^1$\it Department of Physics and Astronomy,
 Indiana University South Bend,}
\centerline{\it South Bend, Indiana, 46634 }
\smallskip
\centerline{${}^{2}$\it Physikalisches Institut, Universit\"at Bonn,
 53115 Bonn}

\vskip .7truein
\centerline{\bf Abstract}
\vskip .1in
\noindent
Theories in more than ten dimensions play an important role in understanding
nonperturbative aspects of string theory.
Consistent compactifications of such theories can be constructed via
Calabi--Yau fourfolds. These models can be analyzed particularly efficiently
in the Landau--Ginzburg phase of the linear $\si$-model, when available.
In the present paper we focus on those $\si$-models which have both a
Landau--Ginzburg
phase and a geometric phase described by hypersurfaces in weighted projective
five-space. We describe some of the pertinent properties of these models,
such as the cohomology, the connectivity of the resulting moduli space, and
mirror symmetry among the 1,100,055 configurations which we have constructed.

\renewcommand\thepage{}
\newpage

\baselineskip=16pt
\parskip=.1truein
\parindent=20pt
\pagenumbering{arabic}

\tableofcontents

\vfill \eject

\section{Introduction}
Over the past years much effort has gone into the exploration of Calabi--Yau
fourfolds.
Such manifolds provide ground states of 4D IIB string theory compactified
on manifolds with positive first Chern class and a nontrivial
dilaton which was
constructed in \cite{cv96, mv96}. The behavior of the dilaton is constrained
by the geometry of the curvature of the threefold and can be summarized
succinctly by considering a fibered fourfold with a section, providing, in a
certain sense, a four-dimensional compactification of a twelve-dimensional
theory,
called F-theory. Such manifolds furnish three-dimensional compactifications of
M-theory and therefore play an important role in the understanding
of certain dualities, such as the N$=$1 4D duality
\cite{svw96, bs96, dgw96, m96, bls96, klry97, m97, ks97, fmw97,
 bjps97, cl97, bcgjl97,bm98}
\beq
\fth(\cyfour) ~~\lolra ~~ \het(V\ra~\cythree),
\lleq{fdual}
where $V\ra~\cythree$ is a stable vector bundle over a Calabi--Yau threefold
determined by the F-theoretic fourfold CY$_4$.

In this context it is of interest to investigate the space of CY fourfolds in
some detail. In the present paper we extend the analysis of Landau--Ginzburg
theories at $c=9$ describing Calabi--Yau threefolds to fourfolds described by
models at $c=12$. We then describe a number of pertinent features
of the resulting moduli space of this class of theories.
The conjectured F-theory/Heterotic duality in 4D leads to a number of
expected properties of the moduli space of CY fourfolds. One natural question
raised by the relation (\ref{fdual}) is the issue of mirror symmetry.
In the context of (2,2) vacua of the heterotic string on Calabi--Yau manifolds
mirror symmetry is known to provide a powerful tool for the analysis of such
ground states. Much less is know about the general framework of (0,2) vacua,
geometrically described by stable vector bundles. As a first step in this
direction the existence of (0,2) mirror symmetry between such vector bundles
was established in \cite{bsw96} for a large class of (0,2) vacua by
generalizing the known (2,2) mirror constructions \cite{gp90,ls90,bh93}. This
analysis shows that different stable bundles indeed correspond to the same
underlying (0,2) superconformal field theory.
More recently conjectures have been put forward which aim at a geometrical
interpretation of (0,2) mirror symmetry \cite{es98,cv98}.
Given mirror symmetry among vector bundles the duality (\ref{fdual}) then
leads to the
expectation that the moduli space of fourfolds is mirror symmetric, at least
in the region described by the subclass of elliptically fibered fourfolds.
As will become apparent however, mirror symmetry in fact holds more
generally, independent of any fibration considerations. Similar to the
case of threefolds we can strengthen the Hodge theoretic mirror symmetry
suggested by the comparison of the cohomology group by a direct construction
of hypersurface mirror pairs via fractional transformations \cite{ls90, ls95}.
For a general D-fold mirror symmetry entails
$h^{p,q}(M) = h^{D-p,q}(M^*)$ for a mirror pair $(M,M^*)$. The Hodge diamond
of a Calabi--Yau fourfold contains only four varying Hodge numbers
\beq
\matrix{ & & & &1 & & & & \cr
 & & &0 & &0 & & & \cr
 & &0 & &h^{1,1} & &0 & & \cr
 &0 & &h^{2,1} & &h^{2,1} & &0 & \cr
 1 & &h^{3,1} & &h^{2,2} & &h^{3,1} & &1 \cr
 }
\eeq
three of which are independent.
A plot of three independent Hodge numbers turns out to be somewhat
unilluminating. As in the case of threefolds however, mirror symmetry for
fourfolds distinguishes the combinations $(h^{1,1}-h^{3,1})$ and
$(h^{1,1}+h^{3,1})$ and therefore we can summarize mirror
symmetry among Calabi--Yau fourfolds in a diagram similar to the mirror plot
of \cite {cls90, ks92, krsk92}. The result for the class of hypersurfaces is
shown in Figure 1.

\vskip .2truein
\par\noindent
 \centerline{\psfig{figure=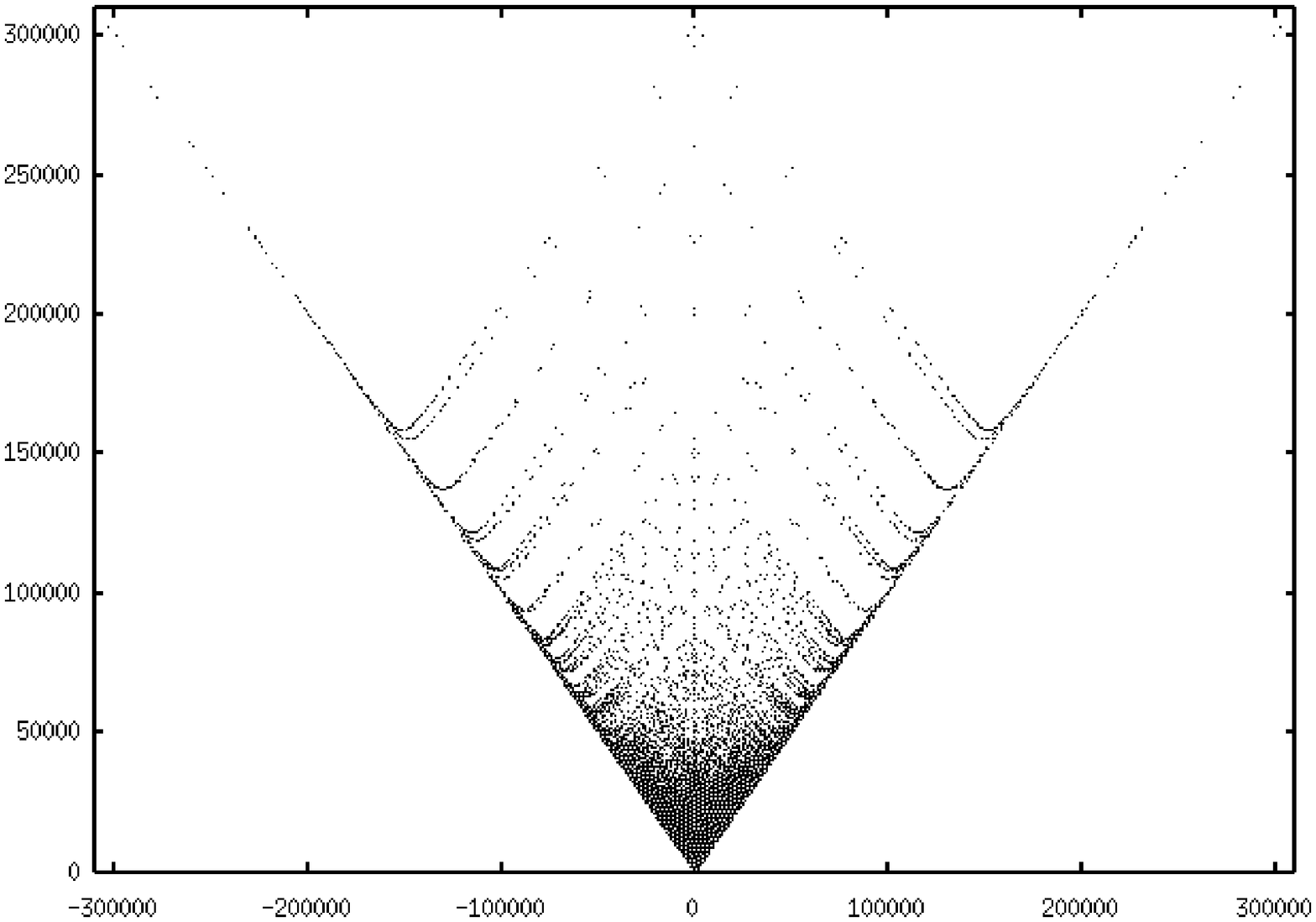,width=15cm,angle=270}}
\par\noindent

\noindent
{\bf Figure 1:}{\it ~Plot of $(h^{(3,1)}+h^{(1,1)})$ vs.\
 $(h^{(3,1)}-h^{(1,1)})$ for the class
 of Calabi--Yau fourfold hypersurfaces in weighted $\IP_5$.}

A further important property of this space is that it is possible to
connect manifolds of this space via certain types of phase transitions
\cite{bls96}, involving singular configurations whose lower dimensional
counterpart are the conifold transitions introduced in \cite{cdls88}.

A number of general aspects, such as the problem of finiteness of the
number of Landau--Ginzburg configurations and transversality of the
allowed potentials are independent of the dimension of the
manifolds. We review our earlier discussion of
these issues for threefolds \cite{ks92} in order to make this
paper self-contained. The article is organized as follows.
After describing in Section 2 and 3 the class of theories we will focus
on, we review in Section 3 the computation of the spectrum of such
theories. In Section 4 we will turn to a discussion of mirror symmetry
and in Section 5 we discuss aspects of the connectedness of the
moduli space of the resulting vacua. We then describe the construction
of the models in more detail in Sections 6 and 7 and summarize our results
in Section 8. Several subclasses of models which are of interest in
the context of F-theory and M-theory are described in Section 9.

\section{$\si$-Models}

The physical theories for which the class of Calabi--Yau fourfolds provides
consistent ground states can be succinctly described via linear $\si$-models.
The starting point of the analysis of \cite{ew93} is a U(1) gauge theory in
N$=$2 superspace, extending the standard Landau--Ginzburg action for the
chiral N$=$2 superfields to
\beq
\cA=\cA_{kin} + \cA_{D,\theta}+ \cA_{kin,\Phi_i} + \cA_{W,\Phi_i}.
\eeq
Consider, in the notation of \cite{ew93}, the gauge invariant field strength
\beq
\cF =\frac{1}{2\sqrt{2}} \{\bcD_+,\cD_-\},
\eeq
the kinetic term of which is given by
\beq
\cA_{kin} = -\frac{1}{4e^2} \int d^2z d^4\theta~ \bcF \cF.
\eeq
There are two possible interactions, the $\theta$ angle term and the
Fayet--Illiopoulos D-term. These can be written as
\beq
\cA_{{\rm D},\theta} = \frac{it}{2\sqrt{2}} \int d^2z d\theta^+d\bth^-~ \cF
 + h.c.
\eeq
where
\beq
t=ir + \frac{\theta}{2\pi}
\eeq
and $r$ is the coefficient of the D-term.

To this are added $N$ chiral superfields with U(1)-charge $k_i \in \IN$.
The kinetic energy of these fields is chosen to be
\beq
\cA_{kin,\Phi_i} = \int d^2z d^4\theta ~ \sum_{i=1}^N \bPhi_i \Phi_i
\eeq
and the superpotential is assumed to be of gauge invariant form
\beq
\cA_{W,\Phi} = -\int d^2z d^2\theta ~W(\Phi_i) - h.c.
\eeq
which is supersymmetric because the $\Phi_i$ are chiral and $W$ is
a holomorphic quasihomogeneous polynomial of the chiral fields.

The constant part of the lowest components of the superfields $\Phi_i$
can be thought of as parametrizing the $n$-dimensional complex space
$\IC_n$, assuming, as in \cite{ew93}, that the K\"ahler metric in the
kinetic term of the $\Phi_i$ should be flat.

The bosonic equations of motion for the auxiliary field $D$ in the
superfield $\cF$ and the auxiliary fields $F_i$ of the chiral
superfields $\Phi_i$ become
\beq
D = -e^2 \left(\sum_i k_i |\phi_i|^2 -r\right)
\eeq
and
\beq
F_i = \frac{\del W}{\del \phi_i}.
\eeq

The bosonic potential that one obtains in terms of the matter fields
$\phi_i$ and the auxiliary fields $D$ and $F_i$ is
\beq
U(\phi_i,\si)=\frac{1}{2e^2} D^2
 + \sum_i \left|\frac{\del W}{\del \phi_i}\right|^2
 + 2|\si|^2 \sum_i k_i^2 |\phi_i|^2.
\eeq

Assuming now that the superpotential takes the form\beq
W(\Phi_i) = \Phi_0 \wtW(\Phi_1,\ldots,\Phi_N),
\eeq
where $\wtW$ is a quasihomogeneous polynomial in the variables
$(\Phi_0,\ldots,\Phi_N)$ with weights $(k_0,\ldots,k_N)$ which is assumed to
be transverse, i.e.\ the equations
\beq
\frac{\del \wtW}{\del \Phi_i} =0
\eeq
can be solved only at the origin.

With this potential the bosonic potential becomes
\beq
U(\phi_i,\si)=\frac{1}{2e^2} D^2 + \left|\wtW\right|^2
 +|\phi_0|^2 \sum_{i=1}^N \left|\frac{\del \wtW}{\del \phi_i}\right|^2
 +2|\si|^2\left(\sum_{i=1}^N k_i^2|\phi_i|^2
 +k_0^2 |\phi_0|^2 \right)
\lleq{bospot}
with
\beq
D = -e^2 \left(\sum_{i=1}^N k_i |\phi_i|^2 -k_0 \phi_0 \bphi_0 - r\right).
\eeq
All terms in (\ref{bospot}) are $\geq 0$. Thus in order to minimize
the potential $U$ one has to minimize $D^2$, which leads to different
results, depending on what the behavior of the variable $r$.

\noindent
The $r\gg0$ phase:
In this case not all $\phi_i$ can be zero. Since the polynomial $\wtW$ is
transverse everywhere except at the origin $\del_i \wtW$ is nonzero for some
$i$. Hence $\phi_0$ and $\si$ must be zero. Thus $D=0$ leads to
\beq
\sum_{i=1}^N k_i \bphi_i \phi_i = r,~~~~\phi_0=0=\si.
\eeq
For $k_i=1$ this simply defines a sphere $S^{2N-1} \subset \IC_N$.
Recalling that one has to mod out the U(1) gauge group and also that the
sphere $S^{2N-1}$ can be Hopf-fibered $S^1 \lra S^{2N-1} \lra \IP_{N-1}$
leads to the condition that the constant bosonic components of $\Phi_i$
parametrize a projective space $\IP_{N-1} = S^{2N-1}/U(1)$.
For $k_i\neq 1$ one arrives at a weighted projective space instead.
Furthermore the vanishing of $\wtW$ leads to a geometry for the space of
ground states described by a hypersurface embedded in $\IP_{N-1}$.

\noindent
The $r\ll0$ phase:
In this case the vanishing of $D$ leads to $\phi_0\neq 0$
and hence the term $|\phi_0|^2 \sum_i |\del_i \wtW|^2$ enforces that
$\phi_i=0$ since this is the only place where the partial derivatives of the
superpotential are allowed to vanish, because
of transversality. This fixes the modulus of $\phi_0$ to be
\beq
k_0 |\phi_0| = -r.
\eeq
Because of the gauge invariance the classical vacuum is in fact unique, modulo
gauge transformations. Expanding around this vacuum leads to
massless fields $\phi_i$
(for $N\geq 3$). To find the potential for these massless fields one has to
integrate out the massive field $\phi_0$. Integrating out $\phi_0$ means
setting $\phi_0$ to its expectation value. Thus the effective
superpotential of the low energy theory is
\beq
\wtW = \sqrt{-k_0 r} W(\phi_i).
\eeq
The factor $\sqrt{-k_0r}$ is inessential since it can be absorbed by
rescaling the $\phi_i$. Since the origin is a multicritical point, this
describes a Landau--Ginzburg orbifold.
It is the Landau--Ginzburg formulation of the theory which lends itself for a
further analysis of a number of aspects of these vacua.

\section{Landau--Ginzburg Theories}

The perhaps simplest way to compute the spectrum of the class of models we
consider is via their Landau--Ginzburg phase.

\subsection{Chiral ring structure}

Using a superspace formulation in terms of the coordinates
$(z,\bz,\th^+,\bth^+,\th^-,\bth^-)$ we can view the Landau--Ginzburg phase
of the linear $\si$-model to be described by the action
\beq
\cA = \int d^2zd^4\th~K(\Phi_i,\bPhi_i) + \int d^2zd^2\th^- ~\wtW(\Phi_i)
 + \int d^2zd^2\th^+ ~\wtW(\bPhi_i)
\eeq
where $K$ is the K\"ahler potential and the superpotential
$\wtW$ is a holomorphic function of the chiral
superfields $\Phi_i$.
The ground states of the bosonic potential are the critical points
of the superpotential of the LG theory and therefore we assume that $W$ has
such critical points. We also require
that these critical points are
isolated since we wish to relate the finite dimensional ring of
monomials associated to such a singularity to
the chiral ring of
physical states in the Landau--Ginzburg theory, in order to construct
the spectrum of the corresponding string vacuum.
The fact that the fermions in the theory should be massless furthermore
leads to the constraint that the critical points are
completely degenerate. Finally, we assume that the Landau--Ginzburg potential
is quasihomogeneous, i.e.\ we can assign to each field
$\Phi_i$ a weight $q_i$ such that for any non-zero complex number
$\l \in \IC^{\star}$
\beq
\wtW(\l^{q_1}\Phi_1,\dots,\l^{q_n}\Phi_n) =\l \wtW(\Phi_1,\dots,\Phi_n).
\eeq
The class of potentials we will focus on thus is comprised
of quasihomogeneous polynomials that have an isolated, completely
degenerate singularity (which we can always shift to the origin).

Mathematically then a so-called catastrophe is associated to each of the
superpotentials $\wtW(\Phi_i)$, obtained by first truncating the
superfield $\Phi_i$ to its lowest bosonic component
$\phi_i(z,\bz)$, and then going to the
field theoretic limit of the string by assuming $\phi_i$ to be constant
$\phi_i=z_i$. Writing the weights as $q_i = k_i/d$, we will denote by
$\IC_{(k_1,k_2,\dots,k_n)}[d]$
the set of all catastrophes described by the zero locus of polynomials
of degree $d$ in variables $z_i$ of weight $k_i$.

The affine varieties described by these polynomials are not compact
and hence it is necessary to implement a projection in order to
compactify these spaces. In Landau--Ginzburg language, this amounts to an
orbifolding of the theory with respect to a discrete group $\ZZ_d$ the
order
of which is the degree of the LG potential \cite{cv89}. The spectrum of the
orbifold theory will contain twisted states which, together with
the monomial ring of the potential, describe the complete spectrum of the
corresponding Calabi--Yau manifold. We will denote the orbifold
of a Landau--Ginzburg theory by
\beq
\IC^{\star}_{(k_1,k_2,\dots,k_n)}[d]
\eeq
and call it a configuration.

In the manifold context we are now interested in complex four-dimensional
K\"ahler manifolds, with vanishing first Chern class. For a general
Landau--Ginzburg theory no unambiguous universal prescription for doing
so has been found, and none can exist \cite{ks92}.
One way to compactify amounts to simply imposing projective
equivalence
\beq
(z_1,\ldots,z_n) \equiv (\l^{k_1} z_1,\ldots,\l^{k_n} z_n)
\lleq{proj}
which embeds the hypersurface described by the zero locus of the
polynomial into a weighted projective space $\IP_{(k_1,k_2,\dots,k_n)}$
with weights $k_i$. The set of hypersurfaces of degree $d$ embedded
in weighted projective space will be denoted by
\beq
\IP_{(k_1,k_2,\dots,k_n)}[d].
\eeq
For a potential with six scaling variables this
construction is completely sufficient in order to pass from the
Landau--Ginzburg theory to a string vacuum, provided
$d=\sum_{i=1}^6 k_i$, which is the condition that these hypersurfaces
have vanishing first Chern class. For more than six variables, however,
this type of compactification does not lead to a string vacuum and
the geometric phase is in fact described by higher codimension manifolds
embedded in products of weighted projective space. A simple example is
furnished by the LG potential in six variables
\beq
W=\Phi_1 \Psi_1^2+\Phi_2 \Psi_2^2+\sum_{i=1}^4 \Phi_i^{12}+
\Phi_5^4
\eeq
which corresponds to the exactly solvable model described by the
tensor product of $N=2$ minimal superconformal theories at the levels
\beq
(22^2 \otimes 10^2\otimes 2)_{D^2\otimes A^3},
\eeq
where the subscripts indicate the affine invariants chosen for the
individual factors\fnote{1}{Further explanations and references can be
found in \cite{ls90b}.}. This theory belongs to the LG configuration
\beq
\IC^{\star}_{(2,11,2,11,2,2,6)}[24]
\lleq{lgform}
whose geometrical phase is described by the weighted complete intersection
Calabi--Yau (CICY) manifold in the configuration
\beq
\matrix{\IP_{(1,1,1,1,3,6)}\cr \IP_{(1,1)}\hfill\cr}
\left [\matrix{1&12\cr 2&0\cr}\right]
\lleq{cyform}
described by the intersection of the zero locus of the two potentials
\bea
p_1 &=& x_1^2y_1+x_2^2y_2 \nonumber \\
p_2 &=& y_1^{12}+y_2^{12}+y_3^{12}+y_4^{12} + y_5^4 +y_6^2.
\eea
Here we have added a trivial factor $\Phi_6^2$ to the potential and
again taken the field theory limit via $\phi_i(z,\bz)=y_i$ and
$\psi_j(z,\bz) = x_j$, where $\phi_i$ and $\psi_j$
are the lowest components of the chiral superfield $\Phi_i$ and
$\Psi_j$. The first
column in the degree matrix (\ref{cyform}) indicates that the first
polynomial is of bidegree (2,1) in the coordinates $(x_i,y_j)$
of the product of the projective line $\IP_1$ and the weighted
projective space $\IP_{(1,1,1,3,6)}$ respectively, whereas the second
column shows that the second polynomial is independent of the
projective line and of degree 12 in the coordinates of the
weighted $\IP_4$.

Even though the assumptions just described \cite{mvw, lvwg}
may seem rather reasonable it is clear that it is not the most
general class of F-theory, M-theory, or string vacua.
Although it provides a rather large set of different models
there are vacua which cannot be described in this framework.
An interesting project for the future would be the complete construction
of hypersurfaces in toric varieties. First steps in this direction
have been taken in \cite{ks97}.

\subsection{Landau--Ginzburg cohomology computation}

The cohomology of Calabi--Yau fourfolds is most efficiently computed via
the Landau--Ginzburg model \cite{bs96} along the lines described in
\cite{cv89}. The simplest part of the computations pertains to the
Euler number which can be obtained via
\beq
\chi = \frac{1}{d} \sum_{l,r = 0}^{d-1}~~
 \prod_{lq_i,rq_i \in \ZZ} \frac{d-k_i}{k_i}.
\eeq

The Landau--Ginzburg construction does better and allows to compute all
the Hodge numbers independently. To do so one constructs a Poincar\'e-type
polynomial for the $l^{\rm th}$ twisted sector
\beq
P_l(t,\bt)= \prod_{lq_i\in \ZZ}
 \left ( \frac{1- (t\bt)^{d-k_i}}{1-(t\bt)^{k_i}}\right)
\lleq{poincare}
which leads to the trace
\beq
Tr_l \left((t\bt)^{dJ_0}\right)
= t^{d\left (k_l +{\small \frac{1}{6}}c_T \right )}
 \bt^{d\left (-Q_l +{\small \frac{1}{6}}c_T \right )}
 \prod_{lq_i\in \ZZ}
 \left ( \frac{1- (t\bt)^{d-k_i}}{1-(t\bt)^{k_i}}\right)
\eeq
with charges
\beq
Q_l = \sum_{lq_i {\footnotesize{\not \in}} \ZZ}
 \left (lq_i - [lq_i] - \frac{1}{2} \right )~
\lleq{ramond-c}
and the central charge of those fields which transform nontrivially
under the twist $l$
\beq
{\small \frac{1}{6}}c_T = \sum_{lq_i \notin \ZZ} \left(\frac{1}{2}-q_i\right).
\eeq
Here $t$ and $\bt$ are formal variables, $d$ is the degree of the
Landau--Ginzburg
potential, the $q_i=k_i/d$ are the normalized weights of the fields and
$[lq_i]$ is the integer part of $lq_i$.
Expanding this polynomial in powers in $t$ and $\bt$ it is possible to read
off
the contributions to the various cohomology groups from the different sectors
of the twisted LG-theory. The (2,1)-forms for example are given by the
number of fields of charge (1,1), i.e.\ the coefficient of $(t\bt)^d$.
In general, the number of $(p,q)$-forms are given by the coefficient of
$t^{(3-p)d}\bt^{qd}$ in the Poincare polynomials summed over all sectors
$l=0,\ldots,d-1$. In more detail,
the basic observation is that in the $l^{th}$ twisted sector
the charges of the states are of the form
\beq
(Q_l,-Q_l) + (r,r).
\eeq
The charges $Q_l$, given by (\ref{ramond-c}),
are the contributions to the charge coming from the twisted fields and $r$ is
any of the charges generated by the subring of those fields that are invariant
under
the $l$-twist. These charges are generated by the Poincar\'e polynomial
(\ref{poincare}) of the invariant fields. Given fields of
integral charge one can generate the cohomology classes:
fields with charges $(p,q) \in \ZZ\times \ZZ$ will generate the
cohomology groups H$^{(r,s)}$ according to the relation
\beq
(p,q) \lolra {\rm H}^{(D-p,q)}
\eeq
where $D$ is, in general, the complex dimension of the manifold.

It was first pointed out in \cite{fh54} that in higher dimensions
the dimensions of the cohomology groups are not necessarily independent,
in contradistinction to surfaces and threefolds. For Calabi--Yau
fourfolds this fact leads to the relation \cite{svw96, klry97}
\beq
44+4h^{(1,1)}+4h^{(3,1)}-2h^{(2,1)} -h^{(2,2)} =0
\lleq{cohorel}
and therefore we only have three independent cohomology dimensions
in the present context.

To illustrate this method we consider the fourfold
 $\IP_{(1,1,2,4,4,4)}[16]$.
The Landau--Ginzburg computation leads to the following
contributions to the cohomology:
\beq
\matrix{ & & & &1 & & & & \cr
 & & &0 & &0 & & & \cr
 & &0 & &3 & &0 & & \cr
 &0 & &3 & &3 & &0 & \cr
 1 & &440 & &1810 & &440 & &1. \cr
 }
\eeq
The Euler number agrees with the previous computations.
The contributions of the various twisted sectors to the individual
Hodge numbers are listed in Table 1.

\begin{footnotesize}
\begin{center}
\begin{tabular}{l l l}
\hline
Cohomology Group &Contributing Twisted Sectors
 $h^{(p,q)}_l=h^{(D-p,D-q)}_{d-l}$
 & Dimension \tabroom \\
\hline
H$^{(1,1)}$ &$h^{(1,1)}_7 + h^{(1,1)}_{11} + h^{(1,1)}_{14}$
 &1+1+1=3 \tabroom \\
H$^{(2,2)}$ &$h^{(2,2)}_0 + h^{(2,2)}_3 + h^{(2,2)}_6
 + h^{(2,2)}_8 + h^{(2,2)}_{10} + h^{(2,2)}_{13}$
 &1787+1+1+19+1+1=1810 \tabroom \\
H$^{(1,3)}$ &$h^{(1,3)}_0 + h^{(1,3)}_{8}$ &439+1=440
 \tabroom \\
H$^{(2,3)}$ &$h^{(2,3)}_4$ &3
 \tabroom \\
\hline
\end{tabular}
\end{center}
\end{footnotesize}

\centerline{{\bf Table 1:}{\it ~~Sector contributions
 to the cohomology {\rm H}$^{p,q}(\IP_{(1,1,2,4,4,4)}[16])$.}}

\subsection{Geometry of fourfolds}

Similar to the case of threefolds weighted projective hypersurface fourfolds
are in general singular and must be resolved. These resolutions correspond to
the twisted contributions in the Landau--Ginzburg formulation of the
previous subsections.
It turns out however that the details of the geometric resolution are quite
different from the resolution of threefolds.

As an example which illustrates this
consider again the the fourfold $\IP_{(1,1,2,4,4,4)}[16]$ as one of the
 simplest spaces with a singular curve (this curve itself is smooth).
With $c_4(\IP_{(1,1,2,4,4,4)}[16])=21,288~h^4$
and the fourfold singularities described by the $\ZZ_2$-singular surface
\bea
\ZZ_2~&:&~~S=\IP_{(1,2,2,2)}[8] \nn \\
 & &~~c_2(S) = 26h^2,~~~\chi_{\rm sing}(S)=26
\eea
and the $\ZZ_4$-singular curve
\bea
\ZZ_4~&:&~~C=\IP_2[4], \nn \\
 & &~~\chi_C=-4,~~~{\rm genus}=3
\eea
one finds the Euler number
\bea
 \chi_4 &=& \frac{21,288 \cdot 16}{2\cdot 4\cdot 4\cdot 4\cdot 4}
 -\frac{1}{2}\left(26-\frac{1}{2}(-4)\right)
 +2\left(26-\frac{1}{2}(-4)\right)
 -\frac{1}{4}(-4) + 4 (-4) \nn \\
 &=& 2688.
\eea
This fourfold is a fibration over the base $\IP_1$ whose generic
quasismooth fiber threefold configuration is given by
$F=\IP_{(1,1,2,2,2)}[8]$. This generic fiber degenerates over
$N=16$ points on the base with singular fibers
$F^{\sharp}=\IP_{(1,2,2,2)}[8]$. We can use this fibration structure to
compute the Euler number independently via the fibration formula
\beq
\chi_{4,{\rm fib}}=\chi(\IP_1 -N)\chi(F) + N\chi(F^{\sharp}).
\eeq
Hence the fibration formula gives
\beq
\chi_{4,{\rm fib}}=(2-16)\cdot (-168) + 16(20+1) = 2688,
\eeq
in agreement with the resolution formula.

We now see that the resolution of fourfolds leads to different
ingredients compared to those of threefolds.
The resolution of the singular curve $C=\IP_2[4]$
introduces in the present
case 3 (2,1)-forms and 3 (1,2)-forms even though this is a
$\ZZ_4$-curve of genus $g=3$. On a threefold its resolution would
have a divisor $D$ which is a fiber bundle over the curve $C$
with fiber $\IP_1 \vee \IP_1 \vee \IP_1$. Instead, on the fourfold
the resolution of this genus 3 curve generates only three (2,1)-forms.
Similar differences appear for the remaining nontrivial cohomology
groups.

\section{Connectedness of Moduli Space}

For a number of reasons we would expect the moduli space of Calabi--Yau
fourfolds to be connected \cite{bls96, bls97}.
First, we can consider fibered fourfolds and consider degenerations in the
fibers. This type of transition is particularly obvious if the fibers are
threefolds because we can then consider conifold transitions, for
instance \cite{cdls88}.
It is also possible however to consider lower-dimensional fibers, such
as K3 surfaces or elliptic surfaces.
Even though the cohomology of the fibers cannot change in such transitions
the resulting transitions of the total space are generically
nontrivial. This generalizes the observations of \cite{ls95} to
fourfolds.

\subsection{Transitions between hypersurfaces in weighted projective spaces}

Simple transitions between hypersurfaces can be constructed via discriminantal
splits of fourfolds \cite{bls97}. An example is provided by the
sextic hypersurface which is connected via such a split
to a hypersurface of degree twelve
\beq
\IP_{(1,1,2,2,2,4)}[12]_{2592} ~~\lolra ~~ \IP_5[6]_{2610}.
\lleq{hypsusplit}
This can be seen by rewriting the lhs
hypersurface as a codimension two complete intersection manifold
\beq
\IP_{(1,1,2,2,2,4)}[12] ~~\sim ~~
\matrix{\IP_{(1,1)}\hfill \cr \IP_{(1,1,1,1,1,2)}\cr}
\left[\matrix{2&0\cr 1&6\cr}\right].
\lleq{fiso}
This can be achieved by first going into the Landau--Ginzburg phase
via the associated linear $\si$-model. We denote this LG configuration
by $\IC^*_{(1,1,2,2,2,4)}[12]$, in which we can consider the Fermat section
of the moduli space for concreteness. Once we are in the LG phase we can
add two mass terms $y_i^2$, $i=1,2$,
without changing the renormalization group fixed
point. This leads to an equivalent description of this model as
\beq
\IC_{(1,1,6,6,2,2,2,4)}[12]~ \ni ~
\left\{ \sum_{i=1}^2\left(x_i^{12}+y_i^2\right)
 + \sum_{j=3}^5x_j^6+x_6^3=0\right\}.
\lleq{lgphase}
This theory however is isomorphic to the orbifold
\beq
\IC_{(1,6,1,6,2,2,2,4)}[12]{\Big /}
\ZZ_2^2:~\left[\matrix{1&1&0&0&0&\cdots &0\cr
 0&0&1&1&0&\cdots &0\cr}\right]
~\sim ~ \IC_{(2,5,2,5,2,2,2,4)}[12]
\lleq{lgisom}
as can be seen with fractional transformations \cite{ls90} and
therefore, by reversing the above steps we see that the geometrical
phase of this LG theory is described by the rhs of equation
(\ref{fiso}).
By applying the discriminantal transition to this codimension
two manifold we finally arrive at the rhs of equation (\ref{hypsusplit}).

\subsection{Connectedness to the space of CICYs}

It has been shown in \cite{bls96} that more generally the space of all
Calabi--Yau
fourfolds embedded in products of ordinary projective space
\beq
\matrix{\IP_{n_1}\cr \IP_{n_2}\cr \vdots \cr \IP_{n_F}\cr}
\left [\matrix{d_1^1&d^1_2&\ldots &d_N^1\cr
 d_1^2&d_2^2&\ldots &d_N^2\cr
 \vdots&\vdots&\ddots &\vdots\cr
 d_1^F&d_2^F&\ldots &d_N^F\cr}\right]
\eeq
is connected via splitting type transitions.
By repeatedly applying $\IP_1$-splits of the type
\beq
X~=~Y[(u+v)~~M]~~\lolra ~~
\matrix{\IP_1\cr Y\hfill \cr}
\left[\matrix{1&1&0\cr u&v&M\cr}\right]~=~X_{\rm split}
\lleq{ponesplit}
to any of the projective factors with $n_i >1$ until all corresponding
$d_a^i=1$ and contracting the $\IP_{n_i}$
via
\beq
X = Y\left[\sum_{a=1}^{n+1}u_a~~~M\right]~~\lolra ~~
\matrix{\IP_n\cr Y \hfill \cr}
\left[
 \matrix{1&1&\cdots &1 &0\cr u_1&u_2&\cdots &u_{n+1} &M\cr}
 \right]~=~X_{\rm split}
\lleq{pnsplit}
it follows that all these manifolds are connected
to the simple configuration
\beq
\matrix{\IP_1\cr \IP_1\cr \IP_1\cr \IP_1\cr \IP_1\cr}
\left [\matrix{2\cr 2\cr 2\cr 2\cr 2\cr}\right]_{1440}
\eeq
with Euler number $\chi=1440$ which can be determined via Cherning.
 From Lefshetz' hyperplane theorem we know that $h^{(1,1)}=5$ and
$h^{(2,1)}=0$. The dimension of H$^{(3,1)}$ for this manifold can be
determined by counting complex deformations with the result $h^{(3,1)}=227$.
 From the Euler number we can then determine that final remaining
Hodge number to obtain the complete Hodge diamond
\beq
\matrix{ & & & &1 & & & & \cr
 & & &0 & &0 & & & \cr
 & &0 & &5 & &0 & & \cr
 &0 & &0 & &0 & &0 & \cr
 1 & &227 & &972 & &227 & &1. \cr
 }
\eeq

After having shown that the space of CICYs is connected it is natural
to ask whether it is non-simply connected. This a more difficult
question to answer. There are however many loop
type transitions which can be constructed within the splitting
construction of \cite{bls96}. An example of such a loop is
provided by the spaces of Figure 2.

\beq
\matrix{ &\matrix{\IP_1\hfill \cr \IP_1 \hfill \cr \IP_{(1,1,2,2,2,4)}\cr}
 \left[\matrix{0&1&0\cr 1&0&0\cr 4&2&6\cr}\right]
 &\lolra
 &\matrix{\IP_1\hfill \cr \IP_1 \hfill \cr \IP_1 \hfill \cr
 \IP_{(1,1,2,2,2,4)}\cr}
 \left[\matrix{1&1&0&0\cr 0&0&1&1\cr 1&0&0&1\cr 2&2&2&6\cr}\right]
 &\lolra
 &\matrix{\IP_1\hfill \cr \IP_1 \hfill \cr \IP_{(1,1,2,2,2,4)}\cr}
 \left[\matrix{0&1&0\cr 1&0&0\cr 2&2&8\cr}\right]
 \cr
 & & & & & \cr
 &\updownarrow
 &
 &
 &
 &\updownarrow
 \cr
 & & & & & \cr
 &\matrix{\IP_1 \hfill \cr \IP_{(1,1,2,2,2,4)}\cr}
 \left[\matrix{1&1\cr 4&8\cr}\right]
 &\lolra
 &\IP_{(1,1,2,2,2,4)}[12]
 &\lolra
 &\matrix{\IP_1 \hfill \cr \IP_{(1,1,2,2,2,4)}\cr}
 \left[\matrix{1&1\cr 2&10\cr}\right]
 \cr
 & & & & & \cr
 &
 &
 &\updownarrow
 &
 & \cr
 & & & & & \cr
 &\matrix{\IP_1\cr \IP_5\cr}\left[\matrix{1&1\cr 1&5\cr}\right]
 &\lolra
 &\IP_4[6]
 &\lolra
 &\matrix{\IP_1\cr \IP_5\cr}\left[\matrix{1&2\cr 1&4\cr}\right]
 \cr
 & & & & & \cr
 &\updownarrow
 &
 &
 &
 &\updownarrow
 \cr
 & & & & & \cr
 &\matrix{\IP_1\cr \IP_1\cr \IP_5\cr}
 \left[\matrix{0&1&1\cr 1&0&1\cr 1&1&4\cr}\right]
 &\lolra
 &\matrix{\IP_1\cr \IP_1\cr \IP_1\cr \IP_5\cr}
 \left[\matrix{1&1&0&0\cr 0&0&1&1\cr 1&0&0&1\cr 1&1&1&3\cr}\right]
 &\lolra
 &\matrix{\IP_1\cr \IP_1\cr \IP_5\cr}
 \left[\matrix{0&1&1\cr 1&0&1\cr 2&1&3\cr}\right]
 \cr
 }
\eeq

\centerline{{\bf Figure 2:}{\it ~~An example of two connected loop
 transitions.}}

The existence of these loops indicates that the moduli spaces of threefolds
and fourfolds and, more generally $n$-folds, is not simply connected. If this
turns out to
be correct then this will have important consequences for
approaches to the connectivity of the moduli space via direct cohomological
methods because it would notably impact the splitting properties of
the long exact sequences which have to be
computed\fnote{2}{We are grateful to Werner Nahm for discussions on these
 matters.}
\cite{wn96}.

\section{Mirror Symmetry}

It is clear from Figure 1 and Figure 2 that the space of Calabi--Yau
fourfold hypersurfaces exhibits a high degree of symmetry.
The total number of 667,954 distinct Hodge diamonds leads to
583,824 distinct pairs of combinations
\beq
(h^{(3,1)}+h^{(1,1)}, h^{(3,1)}-h^{(1,1)})
\eeq
which are shown in Figure 1. The degree of mirror symmetry is roughly
70\% and 1205 of the 202,492 mirror pairs are Hodge theoretically
self-mirror, i.e.\
$h^{(3,1)}=h^{(1,1)}$.

\vskip .2truein
\par\noindent
 \centerline{\psfig{figure=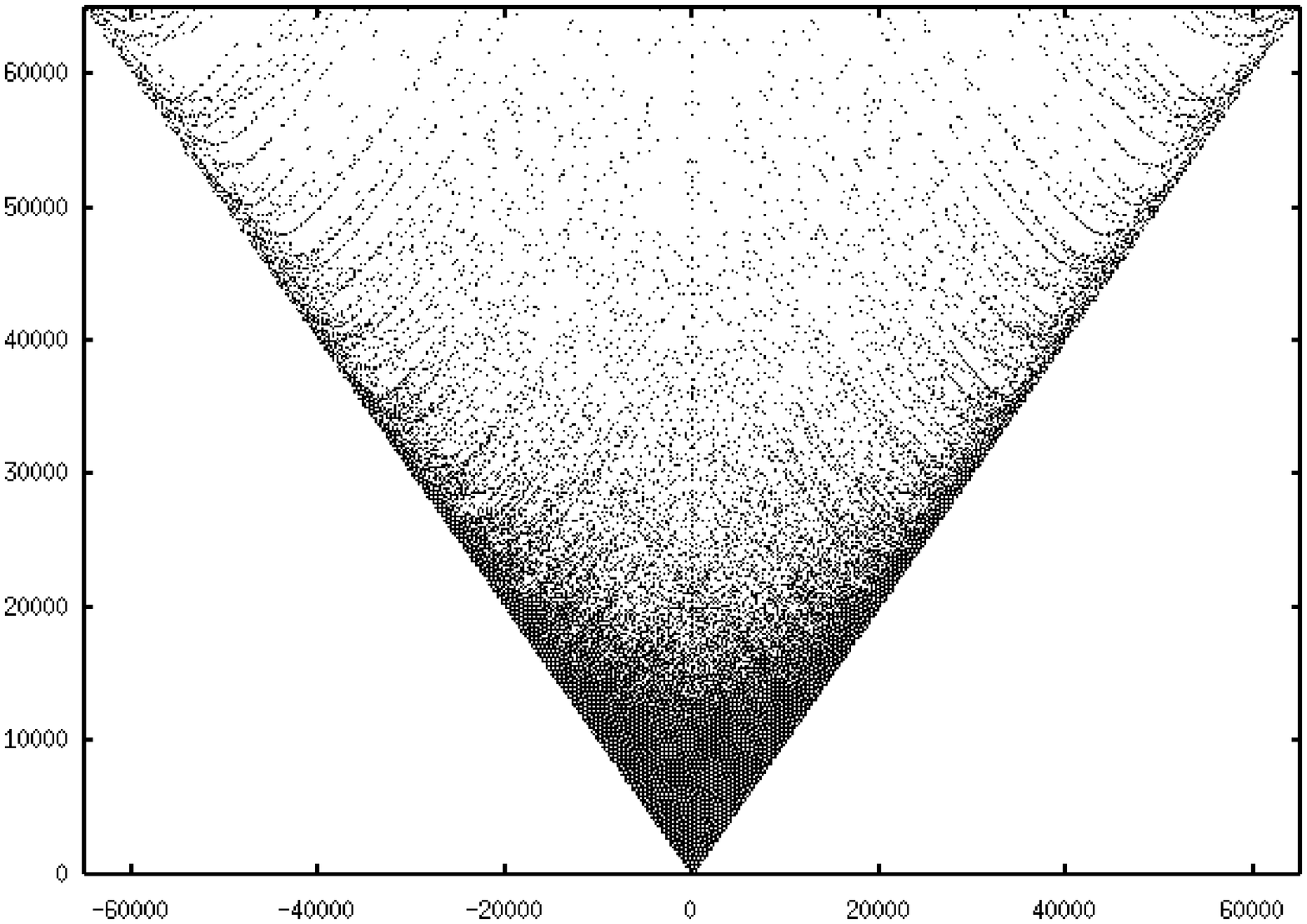,width=15cm,angle=270}}
\par\noindent

\noindent
\centerline{{\bf Figure 3:}{\it ~A zoom of the plot of Figure 1.}}

Similarly to the case of threefold mirror symmetry we
can ask whether potential mirror pairs can be related via
fractional transformations. As expected this is indeed the case.

The essential ingredient of the fractional transformation mirror
construction \cite{ls90,ls95}
is the basic isomorphism
\bea
& &\IC_{\left(\frac{b}{g_{ab}},\frac{a}{g_{ab}}\right)}
\left[\frac{ab}{g_{ab}}\right]
 \ni \left \{z_1^a+z_2^b=0\right \}
 ~{\Big /}~ \ZZ_b: \left[\matrix{(b-1)&1}\right] ~~
 \nn \\ [3ex]
&\sim & \IC_{\left(\frac{b^2}{h_{ab}},\frac{a(b-1)-b}{h_{ab}}\right)}
\left[\frac{ab(b-1)}{h_{ab}}\right]
 \ni \left \{y_1^{a(b-1)/b}+y_1y_2^b=0\right\}
 ~{\Big /}~ \ZZ_{b-1}: \left[\matrix{1&(b-2)}\right]
\llea{basic-iso}

\noindent
induced by the fractional transformations
\bea
z_1 = y_1^{1-\frac{1}{b}}, & & y_1=z_1^{\frac{b}{b-1}} \nn \\
z_2 = y_1^{\frac{1}{b}}y_2, & & y_2=\frac{z_2}{z_1^{\frac{1}{b-1}}}
\eea
in the path integral of the theory.
Here $g_{ab}$ is the greatest common divisor of $a$ and $b$ and
$h_{ab}$ is the greatest common divisor of $b^2$ and $(ab-a-b)$.
The action of a cyclic group $\ZZ_b$ of order $b$ denoted by
$[m~~n]$ indicates that the symmetry acts like
$(z_1,z_2) \mapsto (\a^m z_1, \a^n z_2)$ where $\a$ is the $b^{th}$ root
of unity\fnote{3}{A detailed discussion of the rational structure of the
twisted states implied by the construction of \cite{ls90}
can be found in \cite{ls95}.}.

As an example we consider the simplest fourfold hypersurface configuration
$\IP_5[6]$ described by polynomials of degree six in ordinary projective
5-space. As a first step we consider the action of the cyclic group
of order six defined by
\beq
\ZZ_6 \ni \a :~~(z_1,z_2,z_3,z_4,z_5,z_6)
 \mapsto (\a^5 z_1, \a z_2, z_3, z_4, z_5, z_6),
\eeq
where $\a$ is the sixth root of unity.
Applying the above isomorphism we see that the weighted hypersurface
transform of the orbifold $ \IP_5[6]{\Big /}\ZZ_6:[5~1~0~0~0~0]$
is given by
\beq
\IP_{(6,4,5,5,5,5)}[30] = {\rm Fractional~Transform}\left(
 \IP_5[6]{\Big /}\ZZ_6:~[5~1~0~0~0~0]\right).
\lleq{sextiso}
In principle we have to implement an orbifolding also on the space on the
lhs. But since this $\ZZ_5$ is part of the projective equivalence of the
lhs configuration the relation (\ref{sextiso}) in fact holds.

Applying fractional transformations iteratively to the sextic fourfold
$\IP_5[6]$ with $(h^{(3,1)}, h^{(2,1)}, h^{(1,1)},$ $h^{(2,2)})
=(426,0,1,1752)$
leads to the results in Table 2.
\begin{small}
\begin{center}
\begin{tabular}{|l l l r |}
\hline
Cover Space: $\IP_5[6]$ & & & \tabroom \\
Group &Action &FT Image &($h^{(3,1)}, h^{(2,1)}, h^{(1,1)}, h^{(2,2)})$
 \tabroom \\
\hline
\hline
$\ZZ_6:$ &$[5~1~0~0~0~0]$ &$\IP_{(6,4,5,5,5,5)}[30]$
 &(164,0,6,724) \tabroom \\
$\ZZ_6^2:$ &$\left[\matrix{5&1&0&0&0&0\cr
 0&5&1&0&0&0\cr}\right]$
 &$\IP_{(30,24,21,25,25,25)}[150]$
 &(30,101,30,82) \tabroom \\
$\ZZ_6^3:$ &$\left[\matrix{5&1&0&0&0&0\cr
 0&5&1&0&0&0\cr
 0&0&5&1&0&0\cr}\right]$
 &$\IP_{(150,120,126,104,125,125)}[750]$
 &(6,0,164,724) \tabroom \\
$\ZZ_6^3:$ &$\left[\matrix{5&1&0&0&0&0\cr
 0&5&1&0&0&0\cr
 0&0&5&1&0&0\cr
 0&0&0&5&1&0\cr}\right]$
 &$\IP_{(750,600,630,624,521,625)}[3750]$
 &(1,0,426,1752) \tabroom \\
\hline
\end{tabular}
\end{center}
\end{small}

\centerline{{\bf Table 2:}{\it ~~Fractional transforms of a number of
 group actions on the sextic fourfold.}}

 From this we recognize that the last space is indeed the mirror of the sextic
hypersurface and that also the first space and the third in the table are
mirrors of each other, whereas the second entry is Hodge self-mirror.

\vskip .4truein
\noindent
\section{Transversality of Catastrophes}

\noindent
The most explicit way of constructing a Landau--Ginzburg vacuum is, of
course, to exhibit a specific potential that satisfies all the conditions
imposed by the requirement that it ought to describe a consistent ground state
of the string.
Knowledge of the explicit form of the potential of a LG theory is very
useful information when it comes to the detailed analysis of such a
model. It is however not necessary if only limited knowledge, such as the
computation of the spectrum of the theory, is required. In fact
the only ingredients necessary for the computation of the spectrum
of a LG vacuum \cite{cv89} are the anomalous dimensions of the scaling
fields as well as the fact that in a configuration of weights
there exists a polynomial of appropriate degree with an
isolated singularity. However, it is much easier to check whether
there exists such a polynomial
in a configuration than to actually construct
such a potential. The reason is a theorem by Bertini which asserts that
if a polynomial does have an isolated singularity on the base locus
then,
even though this potential may have worse singularities away from the
base locus, there exists a deformation of the original polynomial that
only
admits an isolated singularity anywhere. Hence we only have to find
criteria
that guarantee at worst an isolated singularity on the base locus.
It is precisely this problem that was addressed in the mathematics
literature \cite{arf89} at the same time as the explicit construction of LG
vacua was started in ref. \cite{cls90}. For the sake of completeness
we briefly review the main point of the argument of ref. \cite{arf89}.

Suppose we
wish to check whether a polynomial in $n$ variables $z_i$ with weights
$k_i$ has an isolated singularity, i.e.\ whether the condition
\beq
dp=\sum_i \frac{\del p}{\del z_i} dz_i = 0
\lleq{transv}
can be solved at the origin $z_1=\cdots = z_n=0$. According to
Bertini's theorem,
the singularities of a general element in $\IC_{(k_1,\ldots,k_n)}[d]$
will lie
on the base locus, i.e., the intersection of the hypersurface and
all the
components of the base locus, described by coordinate planes of
dimension
$k=1,\ldots, n$. Let $\cP_k$ be such a $k$-plane, which we may assume to be
described by setting the coordinates $z_{k+1}=\cdots = z_n$ to zero.
Expand
the polynomials in terms of the non-vanishing coordinates $z_1,\ldots,z_k$
\beq
p(z_1,\ldots,z_n) =
q_0(z_1,\ldots,z_k)~ +~ \sum_{j=k+1}^n q_j(z_1,\ldots,_k)z_j + h.o.t.
\eeq
Clearly, if $q_0\neq 0$ then $\cP_k$ is not part of the base locus
and hence
the hypersurface is transverse. If on the other hand $q_j=0$, then
$\cP_k$ is part of the base locus and singular points
can occur on the intersection of the hypersurfaces defined by
$\cH_j=\{q_j=0\}$.
If, however, we can arrange this intersection to be empty, then the
potential is smooth on the base locus.

Thus we have found that the conditions for transversality in any
number of variables is the existence for any index set
$\cI_k=\{1,\ldots,k\}$ of
\begin{enumerate}
\item{either a monomial $z_1^{a_1}\cdots z_k^{a_k}$ of degree $d$}
\item{or of at least $k$ monomials $z_1^{a_1}\cdots z_k^{a_k}z_{e_i}$
with distinct $e_i$.}
\end{enumerate}

Assume on the other hand that neither of these conditions
holds for
all index sets and let $\cI_k$ be the subset for which they fail. Then
the potential has the form
\beq
p(z_1,\ldots,z_n) = \sum_{j=k+1}^n q_j(z_1,\ldots,z_k)z_j ~+~ \cdots
\eeq
with at most $k-1$ non-vanishing $q_j$. In this case the intersection
of the hypersurfaces $\cH_j$ will be positive and hence the polynomial
$p$ will not be transverse.

As an example for the considerable ease with which one can check whether
a given configuration allows for the existence of a
potential with an isolated singularity, consider the polynomial of Orlik
and Randall
\beq
p=z_1^3+z_1z_2^3+z_1z_3^5+z_4^{45}+z_2^2z_3^4z_4.
\eeq
Condition (\ref{transv}) is equivalent to the system of equations
\beq
\begin{array}{r l r l}
 0 &=~ 3z_1^2+z_2^3+z_3^5, & 0 &=~ 3z_1z_2^2+2z_2z_3^4z_4 \\
 0 &=~ 5z_1z_3^4 + 4z_2^2z_3^3z_4, & 0 &=~ z_2^2z_3^4+45z_4^{44}.
\end{array}
\eeq
which, on the base locus, collapses to the trivial pair of
equations $z_2z_3=0=z_2^3+z_5^5$. Hence this configuration allows for
such a polynomial. To check the system away from the base locus
clearly is much more complicated.

By adding two variables of combined weight 13 it is possible to define
a Calabi--Yau deformation class $\IP_{(1,5,6,8,10,13,15)}[45]$.

\section{Finiteness Considerations}

As in the case of threefolds \cite{ks92, krsk92} the problem of
finiteness has two aspects: first one has to find a constraint on the number
of scaling fields that can appear in the LG theory and then one has to
determine limits on the exponents with which the variables occur in the
superpotential. Both of these constraints follow from the fact that the
central charge of a Landau--Ginzburg theory with fields of charge $q_i$
\beq
c=3\sum_{i=1}^r\left(1-2 q_i\right)=:\sum_{i=1}^r c_i
\lleq{cc}
has to be $c=12$ in order to describe a string, F-theory, or M-theory
vacuum of the type relevant for the structure of the 4D dualities of
interest. The following considerations follow closely the discussion
in \cite{ks92}, adapted suitably to fourfolds.

It follows from the above considerations that we have to assume,
in order to avoid redundant reconstructions of LG theories,
that the central charge $c_i$ of all scaling fields of the potential
should be positive. In order to relate the potentials to manifolds,
we may then add one or several trivial factors or more complicated
theories with zero central charge.

Using the above results we will derive more detailed
finiteness conditions on the number of fields and the size of the
exponents in the superpotential.

\medskip

To get a lower bound on the number of scaling fields, observe
that from (\ref{cc}) written as
\begin{equation}
\sum_{i=1}^r q_i={r\over 2}-{c\over 6}:=\hat c^{(1)}\label{cbed}
\end{equation}
we obtain $r>{c\over3}$ using the positivity of the charges.

Now let $p$ be a polynomial of degree $d$ in $r$ variables. For the
one-element index set $\cI_1$ the conditions (1.) and (2.) for transversality
imply the existence of integers $n_i\geq2$, $1\leq i\leq r$ and of a map
$\sigma:\cI_r \rightarrow \cI_r$ such that for all $i$ one has
\begin{equation}
n_iq_i+q_{\sigma(i)}=1,
\lleq{qs1}
where $\sigma(i)=i$ if condition (1.) and $\sigma(i)\neq i$ if
condition (2.) holds, respectively.

Let us now see how many nontrivial fields can occur at most.
Fields which have charge $q_i\le{1\over3}$ contribute $c_i\ge 1$ to the
conformal anomaly. Next, consider fields with larger charge. Since we assume
$c_i>0$, their charges are in the range ${1\over3}<q_i<{1\over2}$.
It seems that these fields may cause a problem because $c_i\to0$ as
$q_i\to{1\over2}$ which would a priori allow infinitely many fields.
However, among these fields the transversality condition (1.) cannot hold,
because two of them are not enough and three of them are too many
fields in order to form a monomial of charge one. Transversality condition
(2.) implies that a field $z_i$ among them has to occur together with a
partner field $z_{\sigma(i)}$. These pairs contribute $c_i+c_{\sigma(i)}>2$
to the conformal anomaly according to (\ref{cc}) and (\ref{qs1}), so we
can conclude that $r\le c$.

\medskip

In order to construct all transverse LG potentials for a given total
central charge $c$,
we choose a specific $r$ in the range obtained above and consider
all possible maps $\sigma$ of which there are $r^r$. Without restriction
on the generality, we may assume the $n_i$ to be ordered:\
$n_1\le \cdots \le n_r$. Starting with (\ref{cbed}) we obtain via (\ref{qs1})
and the positivity of the charges a bound $n_1<{r\over\hat c^{(1)}}$.

Now we choose $n_1$ in the allowed range and use (\ref{qs1}) in order
to eliminate $q_1$ in favour of the $q_i$, $i>1$. This yields
an equation of the general form
\begin{equation}
\sum_{i=p}^r w^{(p)}_i q_i=\hat c^{(p)}.\label{cbed2}
\end{equation}
In this step we have $p=2$, in (\ref{cbed}) we had $p=1$ and $w^{(p)}_i=1$.
If $\hat c^{(p)}\neq 0$, (\ref{cbed2}) allows us to derive a
finite bound $N_p$ for $n_{p}$:
\beq
n_{p}<{1\over\hat c^{(p)}}\sum_{i\in \cI_\pm} w_i^{(p)}=:N_p,
\lleq{generalbound}
where $\cI_\pm$ are the indices of the positive/negative $w_i^{(p)}$;
the choice depends on the sign of $\hat c^{(p)}$.
If $N_p<n_{p-1}$ we increment $n_{p-1}$ as long as it does not hit its
bound and so on.

What to do in the case $\hat c^{(p)}=0\,$?
If the $w_i^{(p)}$ are indefinite
we get no bound from (\ref{cbed2}). However, we will see that the
existence of monomials for certain index sets, which are required by
the transversality conditions, implies a bound for $n_p$.
Let $\cI_a$ denote the indices of the already bounded
$n_i$ and $\cI_b$ the others. How can indefinite $w$'s arise?
If there is a chain of indices
$a_0=a,a_1=\sigma(a),\ldots,a_l=\sigma^{l-1}(a)=:b(a)$ linked by the map
$\sigma$, the charge of the field $z_a$ with $a\in \cI_a$ will depend on the
unknown charge of a field $z_{b(a)}$ with $b\in \cI_b$.
The charge of $z_a$ is then given by
\beq
q_a={1\over n_a}-{1\over n_a n_{a_1}}+
\cdots -{(-1)^l\over n_a\ldots n_{a_{l-1}}}+
{(-1)^l\over n_a\ldots n_{a_{l-1}}}\ q_{b(a)}.
\lleq{qform}
Indefiniteness of the $w_i^{(p)}$ can only occur if there are fields
$z_a$, $a\in \cI_a$, with odd $l$, i.e.\ the coefficient of $q_{b(a)}$
is negative. Call the index set of these fields $\cI^*$.
Assume first that the transversality condition (1.) holds.
This implies the existence of positive integers $m_i$
such that $\sum_{i\in \cI^*} m_i q_i=1$. From (\ref{cbed}) and the
positivity of the charges it
follows that $m_i<2n_i$. For the unknown $q_i$, $i\in \cI_b$, we get an
equation of the form $\sum_{i\in\cI_b}v_i q_i=\varepsilon$, which yields
a bound for $n_k$ with $k={\rm min}_{a\in\cI^*}b(a)$, since all
$v_i=v_{b(a)}={m_a\over n_a\ldots n_{a_{l-1}}}$ are positive.
The lowest possible positive value $\varepsilon_{\rm min}$ for
\beq
\varepsilon=-1+\sum_{a\in\cI^*}m_a\left({1\over n_a}-\ldots-{(-1)^l\over
n_a\ldots n_{a_{l-1}}}\right).
\lleq{epsexpl}
can be obtained by minimizing $\varepsilon$ with respect to the set
$\{m_i\}$ leading to the bound
\beq
n_i<{2\over\varepsilon_{\rm min}}\sum_{a\in\cI^*}{1\over n_{a_1}\ldots
n_{a_{l-1}}}.
\lleq{complbound}
If there is no such $\varepsilon_{\rm min}$ we have to increment $n_{p-1}$.

If transversality condition (2.) applies, we have $|\cI^*|$ equations
of the form $\sum_{i\in \cI^*} m^{(j)}_i q_i=1-q_{e_j}$ which can be
rewritten as $\sum_{i\in \cI_b} v_i^{(j)} q_i=\varepsilon^{(j)}$.
Deriving a bound is similar to the case discussed above.
Only if all $v_i^{(j)}$ happen to be indefinite and all
$\varepsilon^{(j)}$ are zero we get from this condition.
In \cite{ks92} it has been argued that this cannot occur, again due to
the positivity of the charges.

Finally, the $q_i$ are obtained from the $n_i$ by solving the upper
triangular linear equation system (\ref{cbed2}). The weights $k_i$ and the
degree $d$ are then given by $q_i={k_i\over d}$ with minimal denominators and
numerators.

This procedure of restricting the bound for $n_{p}$, given
$n_1,\dots,n_{p-1}$, for each map $\sigma$ was implemented in a C Code.
It allows all configurations to be found without testing unnecessarily
many combinations of the $n_i$.

In the five and six-variable case we have found 360,346 and 739,709
transverse configurations, respectively. By adding a trivial mass
term $z_6^2$ in the five-variable case, the configurations mentioned
so far lead to four-dimensional Calabi--Yau manifolds described as
hypersurfaces in a five-dimensional weighted projective space
by a one-polynomial constraint.
The list of all these examples, including the Hodge numbers, can be found on
the web \cite{websites}. The complete computer run,
carried out on a cluster of 100 MHz DEC Alpha machines, required a total
amount of CPU time of the order of ten years. In order to avoid integer
overflow the computation is performed most easily by using 64-bit
arithmetics.
The computation of the Hodge numbers on the other hand
needed only a few months of CPU time on an ordinary Pentium PC.
Only for the models with the highest
degrees did we use an Ultra Sparc multi processor machine.

The algorithm can be directly extended to the cases with more variables,
i.e.\ Calabi--Yau manifolds with codimension$>$1.
When dealing with $n$ variables, however, one needs to consider $n^n$ maps
$\sigma$. This number of maps grows much faster than the size
of integer weights of these higher codimension configurations decreases.
Therefore such a run
would need even more computer power than the five/six variable case and
is at present beyond the means of the resources available to us.
In order to get some estimate of the number of expected models we did
an exploratory run which we stopped after 30 years of CPU time.
A brute force run would take the degree $d$, start with the minimal value
and compute all partitions into $n$ numbers of $kd$ ($k$ denotes the
codimension).
For each partition we
check whether the corresponding model is transverse. Then we increment
$d$ and so on. Of course, this algorithm will fail when $d$ reaches
higher values but the generated number of models is large enough
in order to estimate the total number of configurations. Such a brute
force run was performed by Kreuzer and Skarke \cite{ks97}
generating approx.\ 500,000 models with codimension one with
$d\leq4000$. During our run for codimension one we could estimate the total
number
of configuration with the help of their data with the expected result of
more than one million models, very close to our actual result.
We therefore expect that the following scaling argument holds:
Suppose we find $N$ transverse configurations with a brute force run and
$M$ of them are contained in the results of our algorithm which has
computed $K$ configurations, but only with a small part of all maps $\sigma$.
Then there should exist approximately $K\,{N\over M}$ transverse
configurations.

 From this estimate we expect about 800,000 models of codimension two,
about 32,500 models of codimension three and about 40 models of
codimension four, respectively. Our run has found 88\% of them.
To actually perform this computation will require much larger resources
than the ones we have at our disposal.

Concerning the Hodge numbers, the models with higher codimensions give
rise to 98,402=17\% additional spectra. Applying the scaling argument we
expect 112,000 additional spectra, i.e.\ a total of about 700,000.

Finally, we mention that the degree of mirror symmetry does not change
significantly when the higher codimension models are included.
About 28\% of the spectra do not have a
mirror partner, so the degree of mirror symmetry is about 72\%.
A plot of the resulting spectra is very similar to the codimension
one case, hence we do not present it in this paper.

\section{Results}

We have constructed 1,100,055 Landau--Ginzburg theories at $c=12$, where
360,346 configuration exhibit a trivial quadratic factor and
the remaining 739,709 correspond to more general
conformal field theories. This class of models leads to
667,954 different spectra, i.e.\ different Hodge triplets
which determine the complete Hodge diamond
$(h^{(1,1)}, h^{(2,1)},h^{(3,1)},h^{(2,2)})$ of the geometric phase.
The configuration with the maximal degree is given by
\beq
\IP_{(1806, 151662, 931638, 2173882, 3260733)}[6521466],
\eeq
with Hodge numbers $(h^{(2,2)}=1,213,644,\,h^{(3,1)}=252,\,h^{(2,1)}=0,\,
h^{(1,1)}=303,148)$.
Only 204 of the more than one million models lead to a negative Euler
number. The 24 different negative Euler numbers that occur are
collected in Table 3.

\begin{center}
\begin{tabular}{|r r r r r r |}
\hline
$-6$ &$-12$ &$-18$ &$-24$ &$-30$ &$-36$ \tabroom \\
$-42$ &$-48$ &$-60$ &$-66$ &$-72$ &$-84$ \tabroom \\
$-90$ &$-96$ &$-120$ &$-132$ &$-138$ &$-144$ \tabroom \\
$-168$ &$-180$ &$-192$ &$-198$ &$-240$ &$-252$ \tabroom \\
\hline
\end{tabular}
\end{center}

\centerline{{\bf Table 3:}{\it ~~The negative Euler numbers of
 hypersurface fourfolds in weighted $\IP_5$.}}

More details on the spectrum of the individual models can be obtained
from the websites \cite{websites} where we list all the configurations
and their cohomology groups. As mentioned in the cohomology section above
not all four cohomology groups are independent because of the
relation (\ref{cohorel}). Any triplet of cohomology dimensions thus
captures the complete information. In Figure 4 we present a
threedimensional plot where we suppress the cohomology group H$^{(2,2)}$.

\vskip .2truein
\par\noindent
 \centerline{\psfig{figure=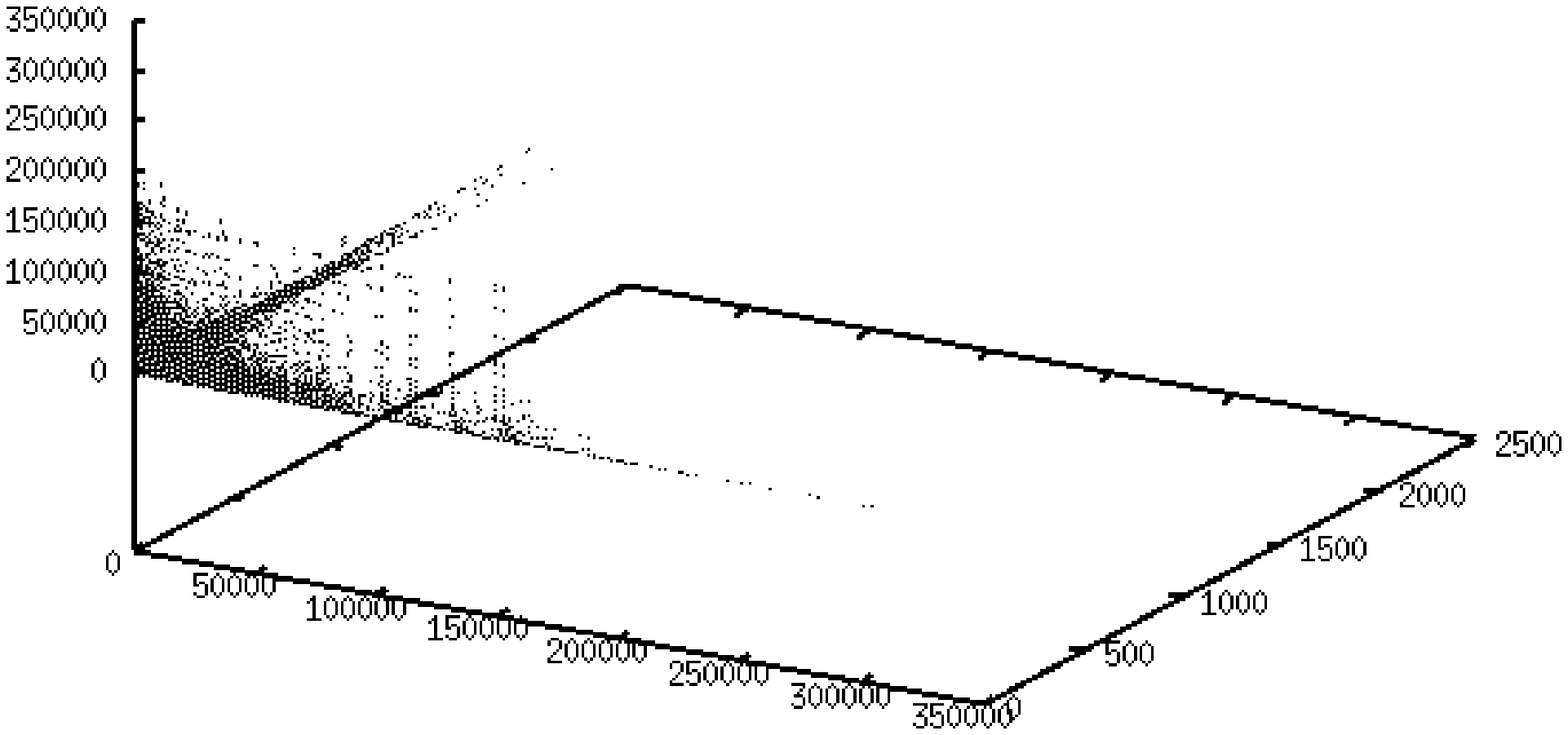,width=15cm,angle=270}}
\par\noindent

\noindent
\centerline{{\bf Figure 4:}
 {\it ~A plot of three independent cohomology dimensions $h^{(1,1)},
 h^{(2,1)}, h^{(3,1)}$.}}

The massless spectrum is very rough information about these models
and as in the case of threefolds Hodge isomorphic theories will
differ in their couplings. As in the case of
threefolds there is some redundancy in our list of spaces.
Consider e.g.\ the two models
\beq
\IP_{(2,2,2,2,1,9)}[18]\ni
 \{z_1^9 + \cdots + z_4^9 + z_5^{18} + z_6^2=0\}
\eeq
and
\beq
\IP_{(1,1,1,1,1,4)}[9]\ni
 \{y_1^9 + \cdots + y_4^9 + y_5^9 + y_5y_6^2=0\}.
\eeq
Using the fractional transform we see that these two theories
are in fact identical, mapped into each other via an orbifolding
which is part of the projective equivalence.

In the fourfold mirror plot of Figure 1 a remarkable parabolic structure
appears in the upper regions of the plot.
In the lower region of Figure 1 this feature
is obscured by the high density of points. The close-up view
of the mirror plot in Figure 3 shows that this feature persists
even for manifolds with smaller cohomology groups.
In Figure 5 we show a close-up view of such curves which suggests
that they do indeed describe parabolas (with a regression coefficient
$r>0.999$).

\par\noindent
 \centerline{\epsfxsize=4.5in\epsfbox{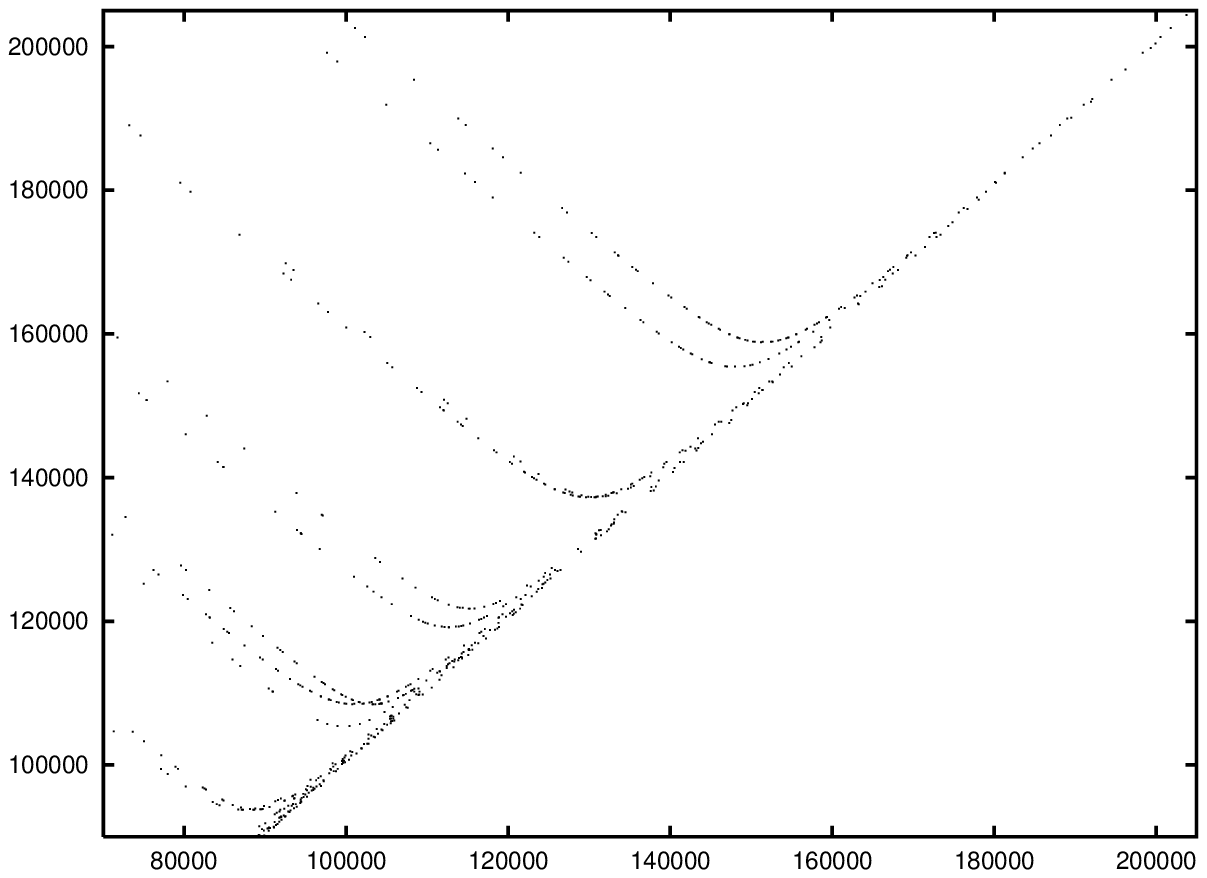}}
\par\noindent

\noindent
\centerline{{\bf Figure 5:}
 {\it ~A close-up view of one of the parabola-like curves.}}

A further zoom in Figure 6 shows that the parabolic structure is not
perfect.

\par\noindent
 \centerline{\epsfxsize=4in\epsfbox{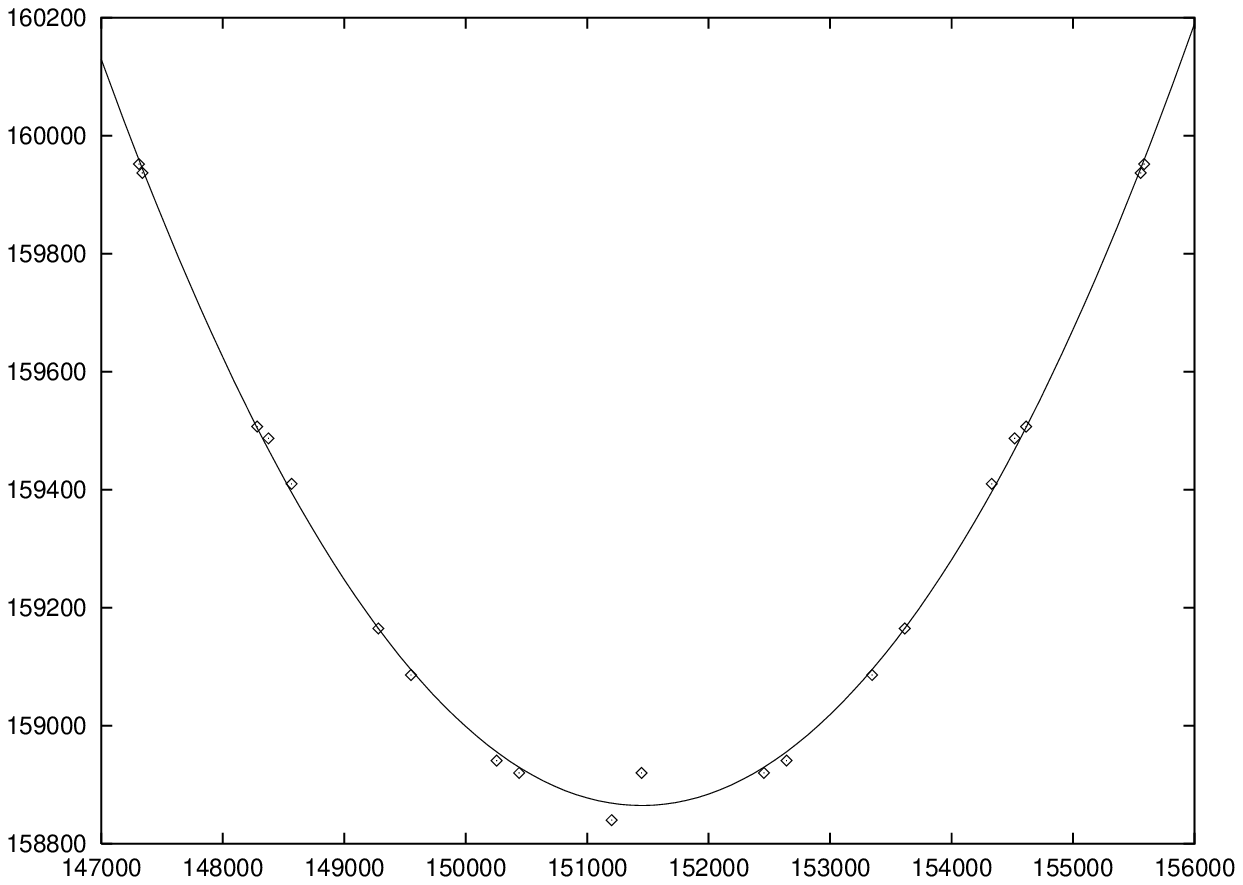}}
\par\noindent

\noindent
\centerline{{\bf Figure 6:}
 {\it ~A close-up view of one of the parabola-like curves.}}

Besides the rigorous relation (\ref{cohorel}) one could
ask for further relations among the Hodge numbers which may only
hold approximately.
Indeed, if we plot $h^{(2,2)}$ vs.\ $(h^{(3,1)}+h^{(1,1)})$ like
displayed in Figure 7, we get the linear relation
\beq
 h^{(2,2)}\approx4.00037\,(h^{(3,1)}+h^{(1,1)})-7.6
\lleq{linrel}
with surprisingly good accuracy, i.e.\ the regression coefficient is
$r=0.999992$.

\vskip .2truein
\par\noindent
 \centerline{\psfig{figure=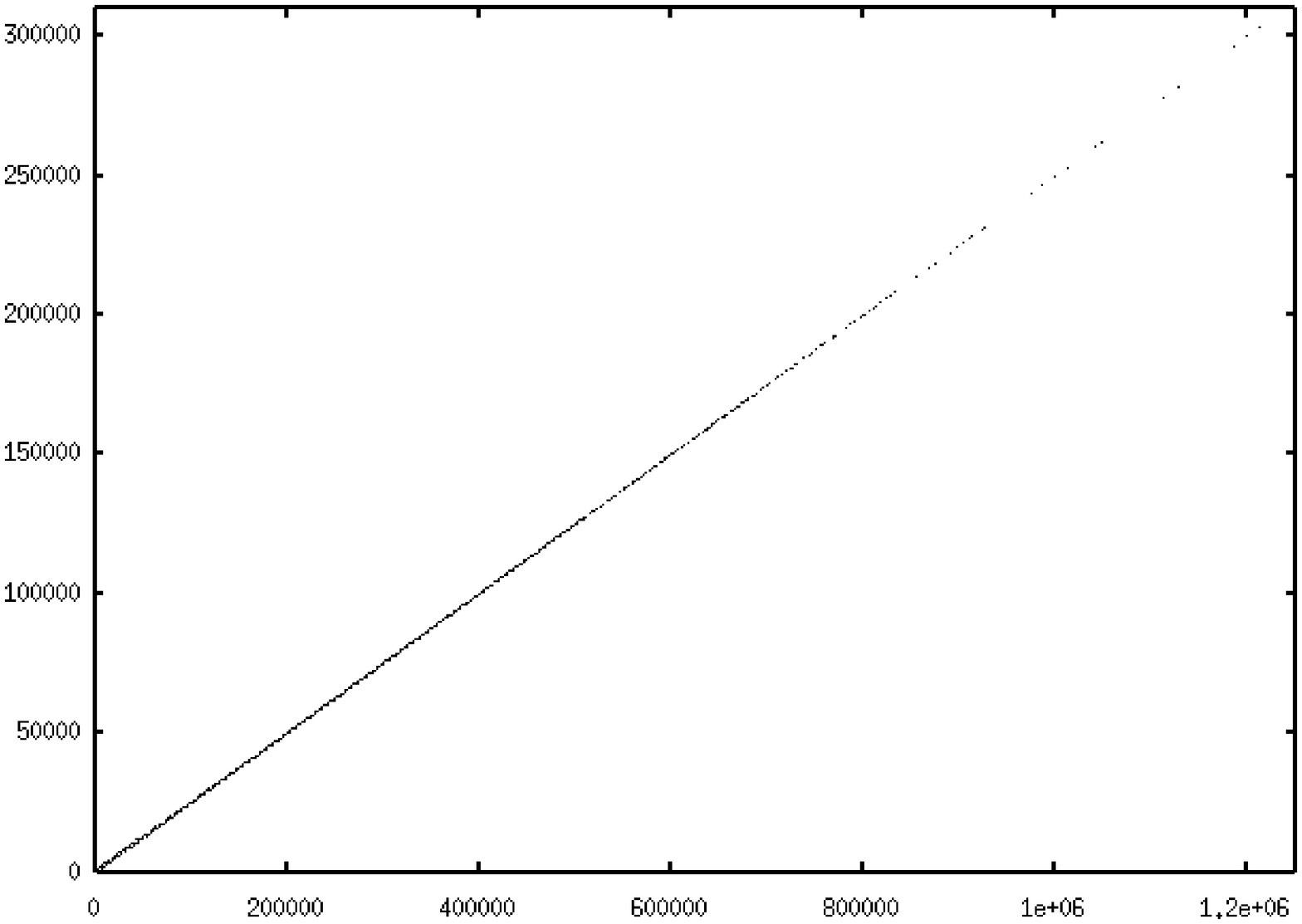,width=12cm,angle=270}}
\par\noindent

\noindent
{\bf Figure 7:}{\it ~Plot of $h^{(2,2)}$ vs.\ $(h^{(3,1)}+h^{(1,1)})$
 for the class of Calabi--Yau fourfold hypersurfaces in weighted $\IP_5$.}

However, (\ref{linrel}) does not hold exactly. To illustrate this, we
show a plot of $(4.00037\,(h^{(3,1)}+h^{(1,1)})-7.6-h^{(2,2)})$ vs.\
log$(h^{(2,2)})$ in Figure 8.

\vskip .2truein
\par\noindent
 \centerline{\psfig{figure=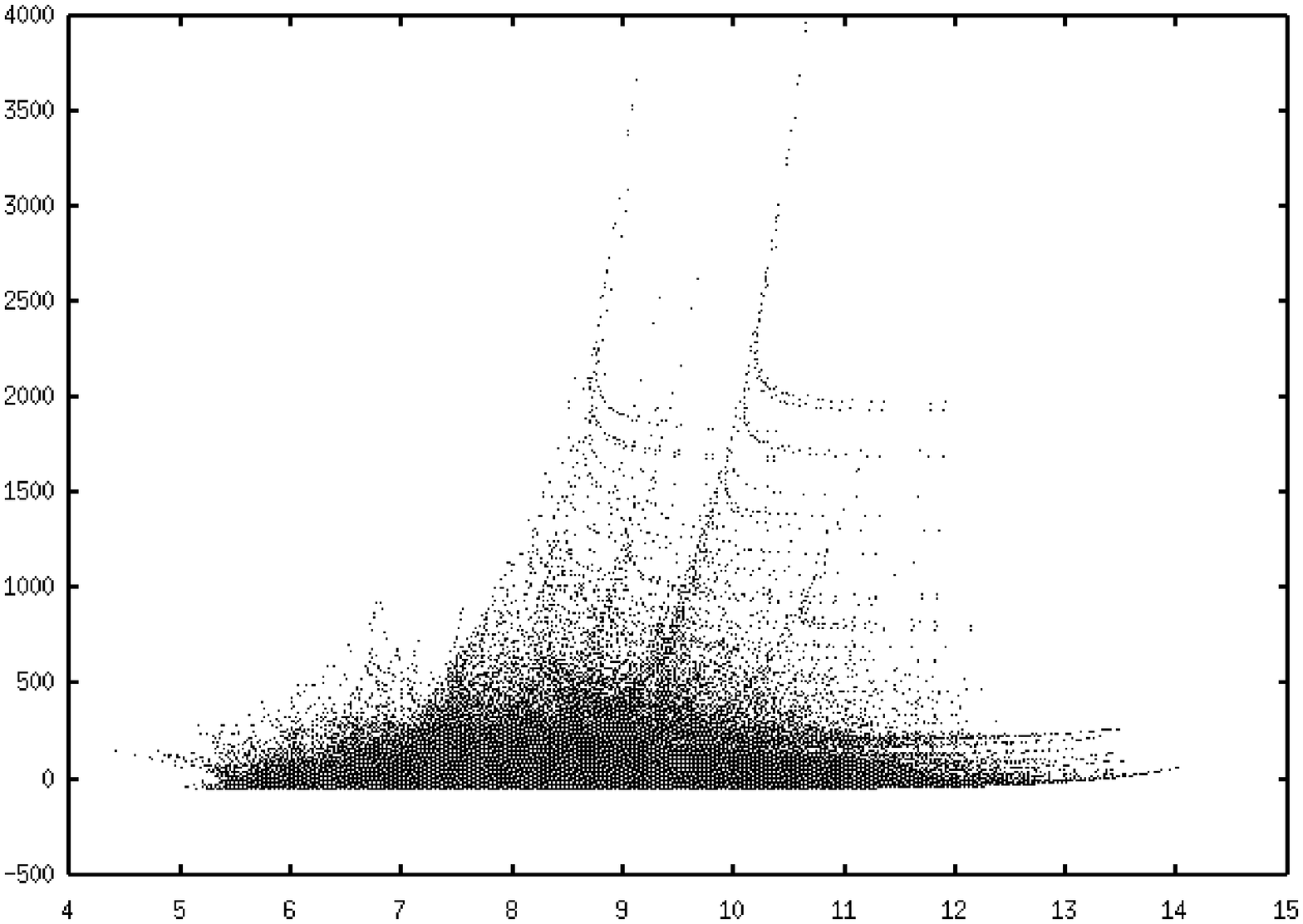,width=12cm,angle=270}}
\par\noindent

\noindent
{\bf Figure 8:}{\it ~Plot of
 $(4.00037\,(h^{(3,1)}+h^{(1,1)})-7.6-h^{(2,2)})$ vs.\ log$(h^{(2,2)})$
 for the class of Calabi--Yau fourfold hypersurfaces in weighted $\IP_5$.}

\section{Fibration Structure}

\noindent
In the context of the recent discussions on duality between F-theory,
M-theory
and heterotic string ground states manifolds that are fibered have become of
particular interest \cite{fibs}.
Some simple sufficient criteria for the existence of fibrations have been
formulated in \cite{hly96}. The idea is to check whether the
$(n-2)$-dimensional subvariety CY$_{n-2}$ defined by a family of divisors
on a Calabi--Yau $(n-1)$-fold
CY$_{n-1}=\IP_{(\tk_1,...,\tk_{n+1})}[\td]$ as
\beq
\cD_i = \{z_i =p_i(z_{j\neq i})\} \cap {\rm CY}_n
\lleq{divhyp}
is itself a Calabi--Yau space. In order to do this a number of conditions
have to be satisfied. If we focus on families of the type
(\ref{divhyp}) we need to be able to partition the weight $k_i$, leading
to the condition (i) $\tk_i=\sum_{j\neq i} b_j\tk_j$.

Furthermore we have to normalize the $(n-1)$-hypersurface appropriately
such that after deleting one variable the remaining ones do not have a
nontrivial common divisor.
To this effect we denote the $n$ remaining weights by
\beq
(k_1,\ldots,k_n)= (\tk_1,..,\hat{\tk_i},...,\tk_{n+1}).
\eeq
The divisor thus defines a hypersurface in $\IP_{(k_1,...,k_n)}[d]$
with $d=\td$ the original degree.
It is this configuration for which it can happen that after deleting a
variable the remaining weights {\it do} have a common denominator, in which
case we would not get a useful configuration. The normalized weight vector
which maps this into a proper configuration is defined as
follows: for each $i$ consider the remaining weights and
denote by $\rho_i$ the greatest common divisor of these weights
\beq
\rho_i := {\rm gcd}(k_1,\ldots,\hat{k}_i,\ldots,k_n).
\eeq
Given these $\rho_i$ we can consider the transformation
\beq
\IP_{(k_1,...,k_n)}[d] \lra \IP_{(\khat_1,...,\khat_n)}[\dhat]
\eeq
defined by the map
\beq
(z_1,...,z_n) \mapsto (x_1,...,x_n):=(z_1^{\rho_1},...,z_n^{\rho_n}).
\eeq
with the weights $\khat_i = \rho_i k_i$. This maps the
the hypersurface defined by the polynomial
\beq
p=\sum_{(a_1,...,a_n),i_j\in\IN }\a_{a_1,...,a_n} z_1^{a_1}\cdots z_n^{a_n}.
\eeq
where $k_j\cdot a_j =d$ into the hypersurface
\beq
\bp = \sum_{(\ba_1,...,\ba_n)}
 \bar{\a}_{\ba_1,...,\ba_n} x_1^{\ba_1}\cdots x_n^{\ba_n}
\eeq
with $\bk_i \cdot \ba_i=\bd$. Hence we need the condition
(ii) $\rho_i | a_i$.

Now, in general the weights
$\khat_i$, $i=1,...,n$ will have a common divisor and in order to obtain
a well-defined configuration we have to divide all weights by this
divisor. To do so consider the set of $\rho_i$'s and define
\beq
\delta_i := {\rm lcm}(\rho_1,\ldots,\hat{\rho}_i,\ldots,\rho_n).
\eeq
Given these, one defines the normalized weight vector
of the final configuration $\IP_{(\bk_1,...,\bk_n)}[\bd]$
by setting $\bk_i = \frac{k_i}{\delta_i}$ and the normalized
degree by $\bd = \frac{d}{\kappa}$ with $\kappa=\rho_i \delta_i$
for any $i$.

With this one can write the original divisor in the threefold whose
normalization defines the fiber as
\beq
\sum_{\bA} \bar{\a}_{\ba_1,...,\ba_n } z_1^{\rho_1 \ba_1}\cdots
                                       z_n^{\rho_n \ba_n}.
\eeq
Since this is a hypersurface of degree $d$ we are interested in a vector $\bA$
such that
\beq
\sum_i \rho_i \ba_i k_i =d.
\eeq
Now, {\it if} the fiber is a Calabi--Yau configuration then we have a
monomial with $\ba_i=1, \forall i$ and hence we have the condition
(iii) $\sum_{i=1}^n k_i \rho_i =d$.

Finally, we can check the transversality of the fiber configuration by
comparing it to the known lists of weighted elliptic curves,
weighted K3 hypersurfaces \cite{arf89} and weighted Calabi--Yau
threefold hypersurfaces \cite{ks92, krsk92}.

By implementing these criteria we can search for a number of different
fibration types among the fourfolds. These criteria are sufficient but
not necessary and hence will not yield an exhaustive list.
They do however provide a wealth of examplest.
The simplest type examples for which
one can analyze problems in F/Heterotic duality are fourfolds which have
an iterative fibration structure in which an elliptic fourfold is
not only a K3-fibration as well but also is CY$_3$-fibered.
The iterative fibrations of such manifolds show a
nested structure which can be summarized in the diagram:
\beq\label{nestfib}
 \matrix{{\rm T}^2 &\lra &{\rm K3} &\lra &{\rm CY}_3
 &\lra &{\rm CY}_4\cr
 & &\downarrow & &\downarrow &
 &\downarrow \cr
 & &\IP_1 & &\IP_1 &
 &\IP_1. \cr}
\eeq
A search for fibrations for which the generic smooth fiber
is a Calabi--Yau threefolds leads to a list 35,540 examples.
Restricting further to such fibrations which are also K3 fibrations
leads to 8,270 examples. These criteria lead to 4,305
examples of this kind which are also elliptic.

For F-theory the threefold fibration structure is not necessary and
we can consider more general fourfolds which are either K3 fibered
or elliptic.
The above criteria lead to a list of 49,751 elliptic fibrations
the Hodges combinations $(h^{(1,1)}\pm h^{(3,1)})$ of which we
display in Figure 9.

\vskip .2truein
\par\noindent
 \centerline{\psfig{figure=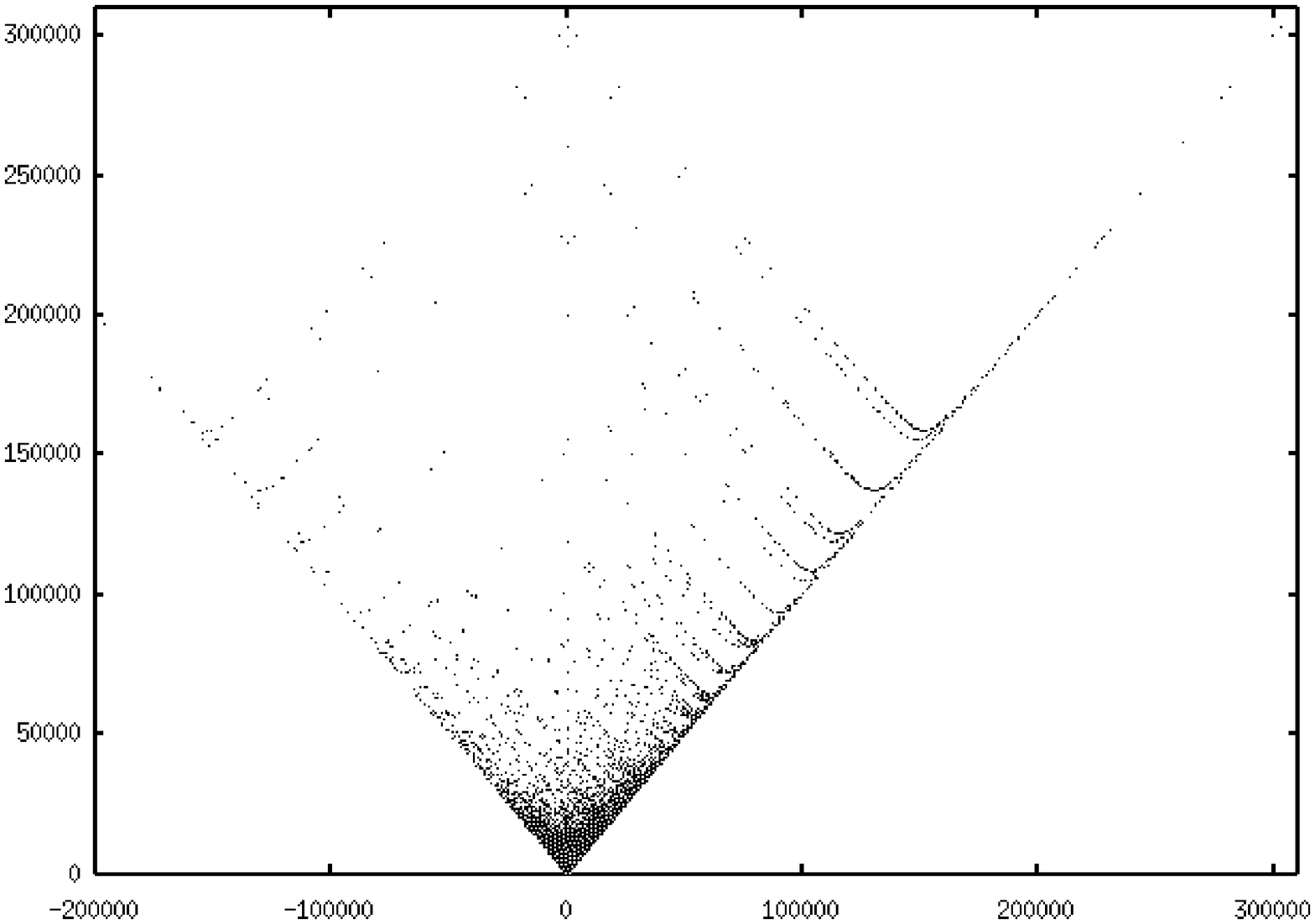,width=15cm,angle=270}}
\par\noindent

\noindent
{\bf Figure 9:}{\it ~Plot of $(h^{(3,1)}+h^{(1,1)})$ vs.\
 $(h^{(3,1)}-h^{(1,1)})$ for 49,751 elliptic
 Calabi--Yau fourfold hypersurfaces in weighted $\IP_5$.}

Of these elliptic fibrations we find that 13,285 pass the test of being
also K3 fibrations. Those spaces we display in Figure 10.

\vskip .2truein
\par\noindent
 \centerline{\psfig{figure=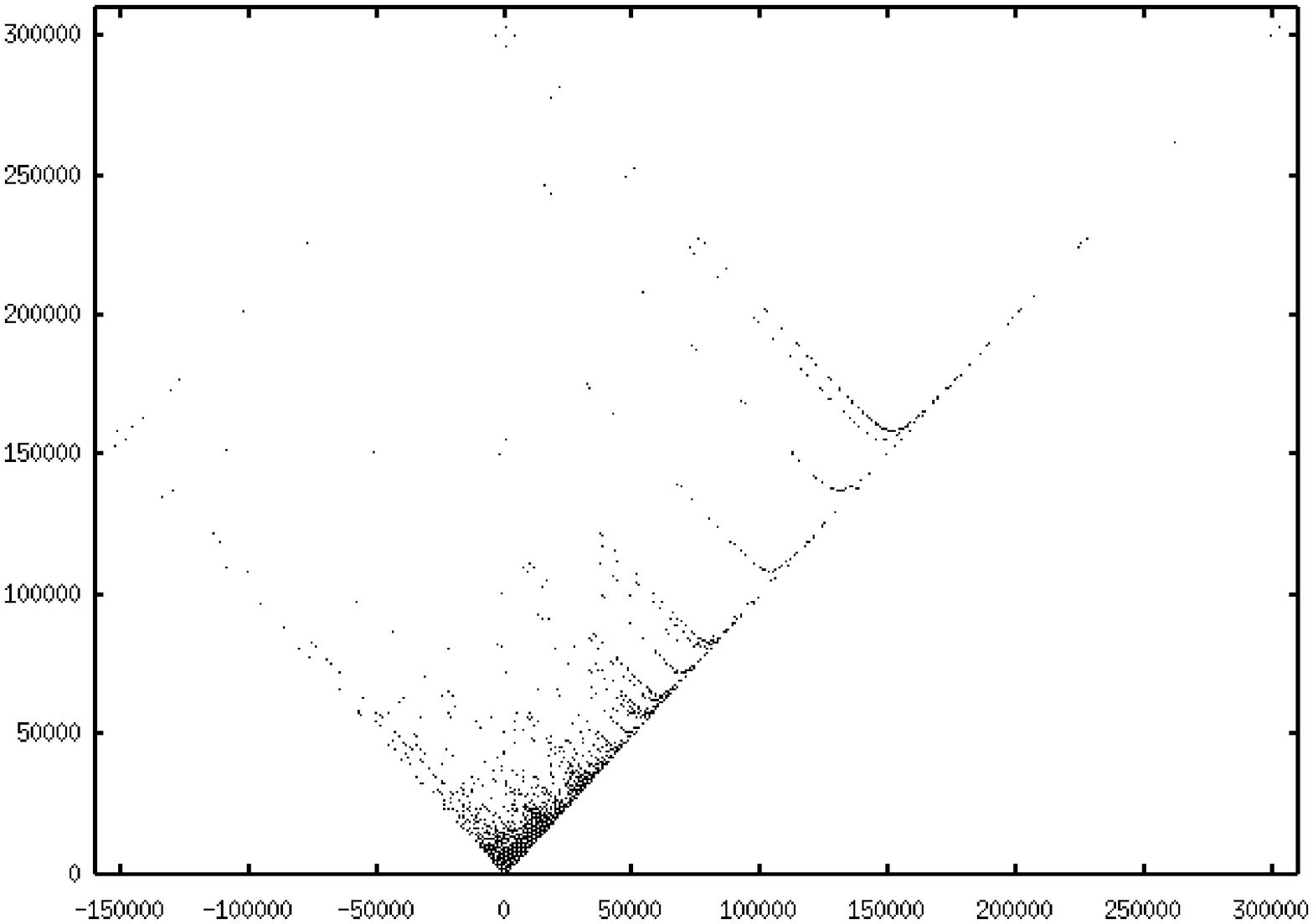,width=15cm,angle=270}}
\par\noindent

\noindent
{\bf Figure 10:}{\it ~Plot of $(h^{(3,1)}+h^{(1,1)})$ vs.\
 $(h^{(3,1)}-h^{(1,1)})$ for 13,285 elliptic
 K3-fibered
 Calabi--Yau fourfold hypersurfaces in weighted $\IP_5$.}

We finally present a plot of the 27099 fourfolds in our list of
Calabi--Yau fourfolds which are K3-fibrations in Figure 11.

\vskip .2truein
\par\noindent
 \centerline{\psfig{figure=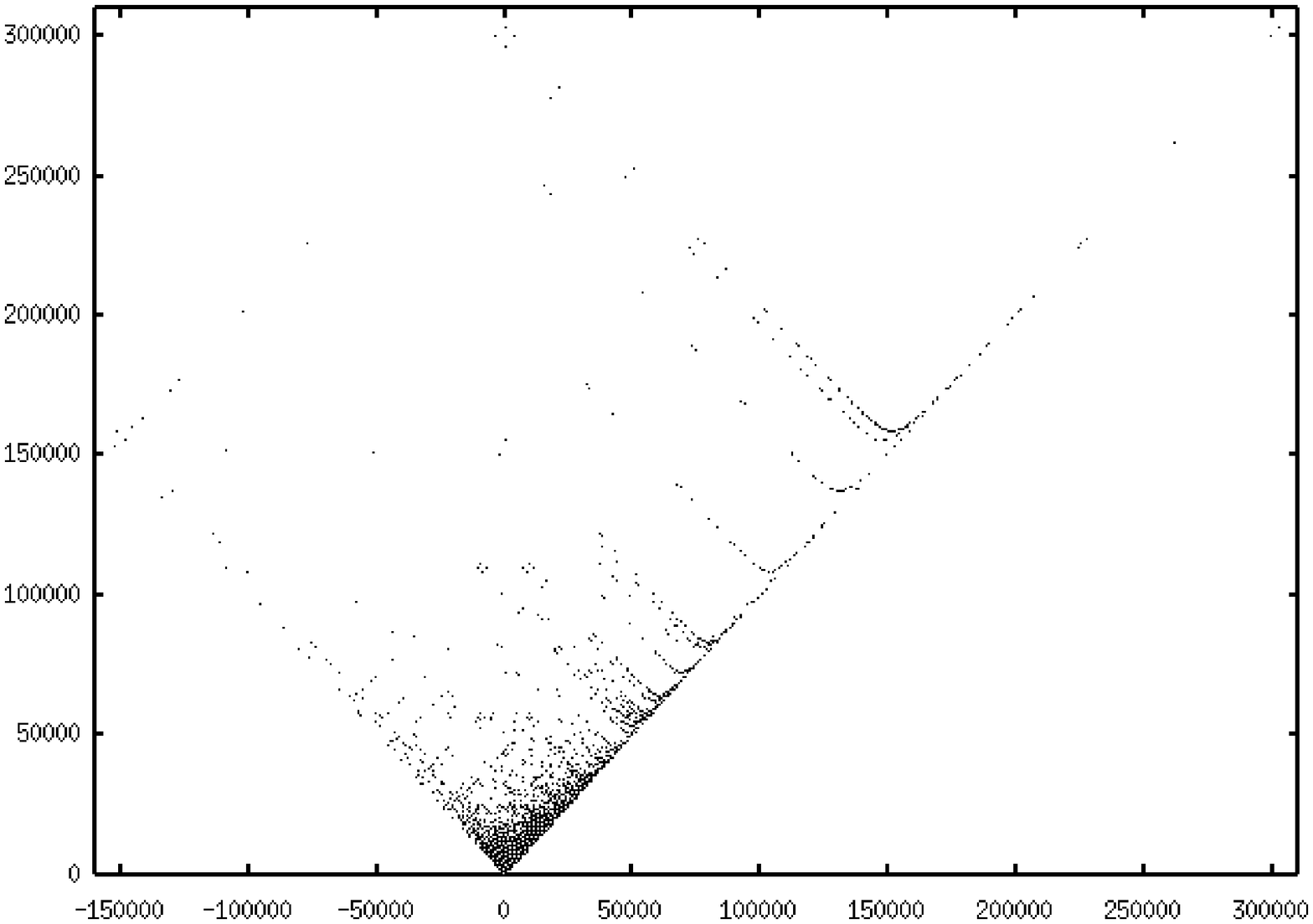,width=15cm,angle=270}}
\par\noindent

\noindent
{\bf Figure 11:}{\it ~Plot of $(h^{(3,1)}+h^{(1,1)})$ vs.\
 $(h^{(3,1)}-h^{(1,1)})$ for 27,099 K3-fibered
 Calabi--Yau fourfold hypersurfaces in weighted $\IP_5$.}

\vskip .3truein
\section*{\bf Acknowledgement}

It is a pleasure to thank Ralph Blumenhagen, Ilka Brunner, and Werner Nahm for
discussions. Part of the computations of this project have been carried out
with the computers of other theoretical physics groups at Bonn
university. We are grateful to Hartmut Monien and Hans-Peter Nilles
for providing these resources.
We also thank the Max Planck Institut for Mathematics at Bonn for providing
computing resources. M.L.\ is grateful to the ITP at Santa Barbara for
hospitality. This research was supported in part by the National Science
Foundation under Grant PHY94-07194.
M.L.\ and R.S.\ are supported in part by NATO, under grant CRG 9710045.

\end{document}